%% file: palatiniMOND.tex
\documentclass[aps,prd,twocolumn,groupedaddress,amsfonts,amssymb,amsmath,showpacs,showkeys,nofootinbib]{revtex4-1}

\usepackage{graphicx}

\usepackage{psfrag}
\usepackage{natbib}

\input{aas-journals}

\begin{document}


\title{A relativistic description of MOND using the Palatini formalism in an 
extended metric theory of gravity}

\author{E. Barrientos}
\email[Email address: ]{ebarrientos@astro.unam.mx}
\author{S. Mendoza}
\email[Email address: ]{sergio@astro.unam.mx}
\affiliation{Instituto de Astronom\'{\i}a, Universidad Nacional
                 Aut\'onoma de M\'exico, AP 70-264, Ciudad de M\'exico 04510,
	         M\'exico \\
            }

\date{\today}

\begin{abstract}
  We construct a relativistic metric description of MOND using the
Palatini formalism following the \( f(\chi)=\chi^b \) description
of \citep{mendozatula}.   We show that in order to recover the
non-relativistic MOND regime where, for circular orbits the Tully-Fisher
law replaces Kepler's third law, the value of the parameter $ b = 3/2 $,
which is coincident with the value found using a pure metric formalism
\citep{mendozatula}.  Unlike this pure metric formalism, which yields
fourth order field equations, the Palatini approach yields second order
field equations, which is a desirable requirement from a theoretical
perspective. Thus, the phenomenology associated to astrophysical phenomena
with Tully-Fisher scalings can be accounted for using this proposal,
without the need to introduce any non-baryonic dark matter particles.
\end{abstract}


\maketitle

\section{Introduction}
\label{introduction}

Newton developed the first successful mathematical theory of
gravitation. This non-relativistic theory of gravity is based on the
empirical foundations of Kepler laws. In particular, Kepler's third law
of motion describes the orbital velocity $v$ of a planet about the sun
as a function of its mass $M$ and separation \( r \)  between 
a given planet:

\begin{equation}
  v \propto \frac{ M^{1/2} }{  r^{1/2} },
\label{kepler}
\end{equation}

\noindent for circular orbits.  Since a particular planet is in centrifugal
equilibrium with the force of gravity, and has a centrifugal acceleration
$a = v^2 / r$, the end result is that the force per unit mass exerted to a
given planet is given by: $a \propto - M / r^2 $.  Newton's idea of gravity
was the result of a mathematical language of forces and accelerations
associated to the empirical observations of Kepler's laws.

  The general relativistic theory of gravity described by Einstein was
built as a wider description of gravitational phenomena embracing 
standard Newtonian gravity as its weak field limit of approximation.  It
has proven extremely well at mass to length ratio scales similar to the
ones associated to our solar system \citep{Will1, Will2, Will3,
Will4,TaylorHulse, Kramer1, Kramer2, Kramer3, Kramer4, Kramer5}.
For these reasons, general relativity  has been taken as the correct
theory to describe gravitation at such scales.

  Observational data of astrophysical systems, including individual, groups
and clusters of galaxies, and the universe in a cosmological context, show
that in order to maintain the standard gravitational field equations of
general relativity, including their Newtonian non-relativistic weak field
limit, it is necessary to postulate the existence a new kind of non-baryonic 
dark matter \citep{Oort, Zwicky, Peebles, Bosma, Rubin, Gunn, Vittorio, 
Gunn-tremaine, Bond-efs, Blumenthal, Neutrinos, Frenk-white}.  Although, current 
research is usually done assuming the existence of this non-detected dark 
matter, the alternative scenario consists on changing the field equations of 
gravitation at those scales.  It was under this point of view
that~\citep{Milgrom1,Milgrom2} developed a MOdified Newtonian Dynamics (MOND) 
approach to non-relativistic gravity.
  
  Shortcomings between the theoretical predictions of general relativity
and astronomical observations have led to propose alternative theories in
order to explain such observations. Amongst these ideas we can  name the
$f(R)$ theories \citep{f(R), liberati, capozziello, Salvatoreextendido,
nojiri11}, Tensor- Vector-Scalar theories \citep{Bekenstein, Bek2,
Sanders-t-v-s, Zlosnik, SkortisTeVeS}, galileons \citep{Babichev},
bimetric theories \citep{Milgrom-bimetric, Bimetric2}, modified energetic
theories \citep{demir2014}, dipolar dark matter \citep{Blanchet1,
Blanchet2, Blanchet3} and nonlocal theories\citep{Soussa, Esposito}.

 In recent years, through dynamical observations of
spiral, elliptical and dwarf spheroidal galaxies \citep{famey,Xavier1},
globular clusters \citep{Xavier2, Xavier3} and even wide open
binaries \citep{Xavier4}, it has became clear that at certain scales of
mass and length, where the induced gravitational accelerations on test
particles are smaller than a certain value \( a_0 \),  Kepler's third
law appears not to hold in its classical form on these systems, but
rather obey the Tully-Fisher law\footnote{To be more precise, the
baryonic Tully-Fisher relation is only observed in spiral galaxies.
For elliptical galaxies an analogous relation is also observed and is 
known as the Faber-Jackson relation~\citep{famey}.  We are using both
relations as to mean the same physical idea, that the scaling with
velocity -or velocity dispersion for pressure supported systems- 
with the mass is the one shown in eq.~\eqref{tully-fisher}.}:

\begin{equation}
  v \propto M^{1/4}.
\label{tully-fisher}
\end{equation}

  Following \citep{mendozareview, sergioyolmo}, we assume that at some
regime, gravity follows Kepler's third law~\eqref{kepler}, and at some
other it follows the Tully-Fisher law.  As such, when the Tully-Fisher
regime is reached, the acceleration exerted by a test particle at
a distance \( r \) from a point mass source \( M \) generating a
gravitational field is given by

\begin{displaymath}
  a = \frac{ v^2 }{ r } \propto - \frac{ M^{1/2} }{ r }.
\end{displaymath}

\noindent As noted by  \citep{mendozareview, sergioyolmo}, the
proportionality constant can be written as \( \sqrt{ G a_0 } \), where 
$a_0 \approx  10^{-10} \, \textrm{m} \,\textrm{s}^{-2} $ is Milgrom's
acceleration constant.  Using this, the previous relation can be written
as:

\begin{equation}
  a = \frac{ \sqrt{ a_0 G M } }{ r }.
\label{mondian}
\end{equation}

  All current observations \citep{Sanders, Mcgaugh, Mcgaugh2, Scarpa,
Mcgaugh3, Sanders2, Milgrom3, Brada-Milgrom} show that
Newtonian gravity is reached when test particles acquire an 
acceleration greater than \( a_0 \) and a full MONDian regime is obtained
when those accelerations are smaller than \( a_0 \).   View in this way,
all systems with accelerations \( a \lesssim a_0 \) are the ones that are
commonly viewed as systems where non-baryonic dark matter is required to
explain the observed dynamics.

  In \citep{mendozatula}, a construction of an extended relativistic
metric theory of gravity that recovers MOND on its weak field limit of
approximation was made.  This construction, has been tested to yield the
correct bending angle of light for gravitational lensing in individual,
groups and clusters of galaxies \citep{mendozalensing} as well as for
the dynamics of clusters of galaxies \citep{sergio-tula-oliver} and for
the accelerated expansion of the universe \citep{mendoza-diego}.

  In this article, we search for a possible extended metric theory of
gravity using the Palatini formalism, which recovers MOND on its weak
field limit of approximation.  In sect.~\ref{theory} we briefly introduce
the relevant equations for the Palatini metric formalism useful for our
further developments.  In sect.~\ref{power} we propose a power of the
Ricci scalar for the gravitational action and we find an expression for
the Ricci scalar curvature as function of the trace of the energy-momentum
tensor. In sect.~\ref{perturbacion} we explore the non-relativistic
weak-field limit of the theory and expand the metric as Minkowskian
plus a second order perturbation to arrive at a non-relativistic
equation for the acceleration as function of the energy-momentum
tensor. In sect.~\ref{mond} we fix the free parameters of our theory
such that in the weak-field limit of approximation the acceleration
converges to the simplest MONDian description of eq.~\eqref{mondian}.
In sect. \ref{PPN} we perform a Parametrised Post Newtonian (PPN)
analysis to second order of our field equations in order to complement
the results of sect.~\ref{perturbacion}.  In sect.~\ref{Conclusiones}
we conclude and discuss our results.

\section{$f(\chi)$ in Palatini formalism}
\label{theory}

Many of the results
mentioned in this section are well known on the studies of the Palatini
formalism for metric \( F(R) \) theories of gravity.  For further
information, the reader is referred to the excellent introductory texts by
\citep{f(R), liberati, capozziello, Salvatoreextendido, nojiri11}.

Let us start with a dimensionally correct action for the gravitational
field motivated by the one built by
\citep{mendozatula}:

\begin{equation}
  \begin{split}
 {\cal{S}} &= - \frac{ c^3 }{ 16 \pi G L_M^2 } \int f(\chi) \sqrt{-g} \, 
   \mathrm{d}^4x \\
  & - \frac{1}{c} \int {\cal{L_\textrm{matt}}} \sqrt{-g} \, \mathrm{d}^4x,
  \end{split} 
\label{action}
\end{equation}

\noindent where \( L_M \) is a coupling constant with dimensions of 
length and the dimensionless Ricci scalar \( \chi \) is given by:

\begin{equation}
  \chi:= L_M^2 \cal{R},
\label{chi}
\end{equation}

\noindent where \( \cal{R} \) is a non-traditional Ricci scalar, not to
be confused with the standard Levi-Civita Ricci's one \( R \).  Both are
related to each other by the following relation:

\begin{equation}
  {\cal{R}} := g^{\mu\nu}  \cal{R}_{\mu\nu}.
\label{RicciPalatini}
\end{equation}

\noindent In the previous equation, and in what follows, we use Einstein's
summation convention, greek and latin indices take values from \( 0  \)
to \( 4 \) and from \( 0 \) to \( 3 \) respectively.  The tensor $g_{\mu\nu}$
represents the metric tensor  and $\cal{R}_{\mu\nu}$ is a non-traditional
Ricci tensor defined exclusively in terms of the affine connection $
^*\Gamma^\alpha\,_{\mu\nu}$ through the following equation:

\begin{equation}
 {\cal{R}}_{\mu\nu} := ^*\Gamma_{\mu\nu,\lambda}^\lambda
-^*\Gamma_{\mu\lambda,\nu}^\lambda
+^*\Gamma_{\mu\nu}^\rho \ ^*\Gamma_{\lambda\rho}^\lambda
-^*\Gamma_{\mu\lambda}^\rho \ ^*\Gamma_{\nu\rho}^\lambda. 
\label{definicionricci}
\end{equation}

\noindent  In the Palatini formalism, the connection
$^*\Gamma^\alpha\,_{\mu\nu}$ has no relation with the standard Levi-Civita
connection $\Gamma^\alpha\,_{\mu\nu}$.

  The null variations of the action~(\ref{action}) with respect to the
metric \( g_{\mu\nu} \) yield the following field equations:

\begin{equation}
  f'(\chi)\chi_{\mu\nu}-\frac{1}{2}f(\chi)g_{\mu\nu} = 
    \frac{8\pi G L_M^2}{c^4}T_{\mu\nu}, 
\label{campo1}
\end{equation}

\noindent where the dimensionless tensor

\begin{equation} 
  \chi_{\mu\nu} := L_M^2 \cal{R}_{\mu\nu} 
\label{tensorchi}
\end{equation}

\noindent and $ f'(\chi) := \mathrm{d} f(\chi)/ \mathrm{d}
\chi $. The energy-momentum tensor \( T_{\mu\nu} \) is given
by~\citep{gravitation}:

\begin{equation}
  T_{\mu\nu} :=
    -\frac{2}{\sqrt{-g}}\frac{\delta({\cal{L}}_\text{matt}\sqrt{-g})}{\delta
    g^{\mu\nu}}.
\label{variacionmateria}
\end{equation}

The contraction of eq. (\ref{campo1}) with  $g^{\mu\nu}$ yields:

\begin{equation}
  L_M^2f'(\chi){\cal{R}}-2f(\chi)=\frac{8\pi G L_M^2}{c^4}T,
\label{trazafinal}
\end{equation}

\noindent for all \( f(\chi) \neq \chi^2 \).  Under the assumption
of a torsion free connection, i.e. imposing a symmetric connection
$^*\Gamma^\alpha\,_{\mu\nu}$, the null variations of the
action~\eqref{action} with respect to this affine connection yield:

\begin{equation}
  \nabla_\lambda \left( \sqrt{-g} \, f'(\chi) \, g^{\mu\nu} \right) = 0.
\label{campo2}
\end{equation}

  The usual approach to solve this equation, consists on performing the
following conformal transformation to the metric tensor:

\begin{equation}
  h_{\mu\nu} = f'(\chi) g_{\mu\nu}.
\label{conforme}
\end{equation}

\noindent Substitution of this last equation into relation~(\ref{campo2})
gives:

\begin{equation}
  \nabla_\lambda\left(\sqrt{-h}h^{\mu\nu}\right)=0,
\label{campo3}
\end{equation}

\noindent where $h:=h^\nu\,_\nu$.  Equation~\eqref{campo3} is known as the
metricity condition and states that $^*\Gamma^\alpha\,_{\mu\nu}$
is the Levi-Civita connection of the $h_{\mu\nu}$ metric, i.e.:

\begin{equation}
  ^*\Gamma_{\mu\nu}^\lambda=\frac{1}{2}h^{\lambda\rho}\left(h_{\rho\mu,\nu}
    +h_{\rho\nu,\mu}-h_{\mu\nu,\rho}\right). 
\label{levicivita}
\end{equation}

  For the conformal transformation~\eqref{conforme}, the tensor
${\cal{R}}_{\mu\nu}(^*\Gamma)$ is related to the usual Ricci tensor
$R_{\mu\nu}(\Gamma)$ defined in terms of the Levi-Civita connection of the
metric $g_{\mu\nu}$  by \citep{carroll, capozziello}:

\begin{equation}
 \begin{split}
   {\cal{R}}_{\mu\nu} =& R_{\mu\nu}-\frac{1}{f'}\nabla_\mu\nabla_\nu f'
      -\frac{1}{2f'}g_{\mu\nu}\Delta f'  \\
   +& \frac{3}{2f'^2}\nabla_\mu f'\nabla_\nu f', 
 \end{split}
\label{conforme1}
\end{equation}

\noindent The contraction of this last result with the metric $g^{\mu\nu}$
yields:

\begin{equation}
  {\cal{R}}=R-\frac{3}{f'}\Delta f'+\frac{3}{2f'^2}\nabla_\mu f'\nabla^\mu f'.
\label{conforme2}
\end{equation}

\noindent  Note that ${\cal{R}}$ is not the Ricci scalar for the $h_{\mu\nu}$
metric, since it is built by its contraction with the conformal
metric $h^{\mu\nu}$.

  In what follows we are going to work extensively with 
eqs.~(\ref{trazafinal}) and~(\ref{conforme2}), since 
using the former it is possible to find 
$\cal{R}$ as a function of the trace of the energy-momentum tensor, 
i.e.: $\cal{R}=\cal{R}(T)$.  Substitution of this result on  
the latter, and bearing in mind the fact that $f'$ is a function of
$\cal{R}$ and hence of $T$, the solution $R=R(T)$ can be found.

\section{$f(\chi)$ as a power law}
\label{power}

  Let us now assume that:
 
\begin{equation}
  f(\chi)=\chi^b. 
\label{potencia}
\end{equation}

\noindent and substitute it into relations~(\ref{conforme2})
and~(\ref{trazafinal}) to obtain:

\begin{gather}
 {\cal{R}} = R+\frac{3}{2}{\cal{R}}^{2-2b}\nabla_\mu {\cal{R}}^{b-1}\nabla^\mu {\cal{R}}
^{b-1}-3{\cal{R}}^{1-b}\Delta {\cal{R}}^{b-1},  
		\label{R2} \\
\intertext{and}
  {\cal{R}}^b=\frac{\alpha T}{b-2}, 
			\label{traza1} \\
\intertext{where}
  \alpha := \frac{8\pi GL_M^{2(1-b)}}{c^4}.
  			\label{alpha} 
\end{gather}

\noindent In order to obtain an equation that relates the 
curvature $R$ with the trace \( T \) of the energy-momentum tensor,
eq.~\eqref{traza1} must be substituted into \eqref{R2}. 
Since this procedure yields a complex equation, we 
will tackle the problem in a different manner.

  Let us then proceed by expressing relation \eqref{conforme1} as:

\begin{equation}
	{\cal{R}}_{\mu\nu} = R_{\mu\nu} + H_{\mu\nu}({\cal{R}}),
	\label{Taylor1}	
\end{equation}

\noindent where we have used the fact that $f'=f'({\cal{R}})$. The tensor
\( H_{\mu\nu}({\cal{R}}) \) has a complicated algebraic form which will be
determined in sect.~\ref{PPN},
and we will show  an explicit functional form which
allows our proposal to have full consistency at the lowest
second perturbation order. The trace of eq.~\eqref{Taylor1} is:

\begin{equation}
	{\cal{R}} = R+ H({\cal{R}}),
	\label{Taylor2}	
\end{equation}

\noindent which is another way to express 
eq.~\eqref{conforme2}.  A Taylor expansion of the function 
$H({\cal{R}})$ yields the following linear relation: 

\begin{equation}
	 H({\cal{R}})= \kappa {\cal{R}} + \mathcal{O}({\cal{R}}^2),
	\label{lineal}
\end{equation}

\noindent since \( H(R=0)=0 \) according to eq.~\eqref{Taylor2}. 
Substitution of eq.~\eqref{lineal} into~eq.~\eqref{Taylor2} yields:

\begin{equation}
	R=\kappa'{\cal{R}}, \qquad \text{where:} \qquad \kappa':=1-\kappa.
	\label{Rlineal}
\end{equation}

Using this result in eq.~\eqref{traza1}, we obtain:

\begin{equation}
  R = \kappa'\left[\frac{\alpha T}{b-2}\right]^{1/b}. \label{Rfinal}
\end{equation}

\section{Weak field limit}
\label{perturbacion}

Our main target is to find $b$ such that in the weakest (non-relativistic)
limit of the theory, the acceleration of a test particle in a
gravitational field produced by a point mass source \( M \)  is reduced to the
MONDian one~\eqref{mondian}.

For this purpose we take the background metric as the Minkowsky space-time 
plus a small perturbation expanded in powers of $1/c$, which we call
perturbation orders.  As an example, a second perturbation order is proportional
to \( 1 / c^2 \) and zeroth order terms have no dependence on the speed of
light.  The next perturbation expansion to the Minkowsky background metric
is of second order \citep{Will}, and since we are interested in the weakest
limit of the theory describing the motion of non-relativistic massive
test particles, this correction is enough for our study. 

At this point we stress that in eq.~\eqref{Rfinal} we have returned to
the original metric $g_{\mu\nu}$. The conformal transformation was
just a mathematical tool in order to manipulate more easily the resulting
equations. Therefore, the expansion used below is justified.

For the second perturbation order, we take as base the work of
\citep*{sergioyolmo}, in which they proved that, to be in accordance
with astronomical observations of the deflection of light of individual,
groups and clusters of galaxies together with the Tully-Fisher
law~\eqref{tully-fisher} for material particles, the metric coefficients 
at second perturbation order are given by:

\begin{equation}
  \begin{split} 
  g_{00} &= {}^{(0)}g_{00} + {}^{(2)}g_{00}=1+\frac{2\phi}{c^2},\\
  g_{ij} &= {}^{(0)}g_{ij} + {}^{(2)}g_{ij}=
  \delta_{ij}\left(-1+\frac{2\phi}{c^2}\right), \\
  g_{0i} &= 0. 
  \end{split}
\label{segundoorden}
\end{equation}

\noindent which implies that the PPN  parameter $\gamma=1$ according
to \cite{sergioyolmo}.  In the previous equation and in what follows
the left superscript in parenthesis on a given quantity denotes its 
perturbation order.  Ricci's scalar of the previous metric at the same
perturbation order is given by:

\begin{equation}
  {}^{(2)}R=-\frac{2\nabla^2\phi}{c^2}. 
	\label{Rperturbado}
\end{equation}

  Since the Tully-Fisher law describes the motion of non-relativistic dust
particles, then the energy-momentum tensor trace is

\begin{equation}
  T=\rho c^2,
\label{polvo}
\end{equation}

\noindent where $\rho$ is the mass density. Thus, eq.~(\ref{Rfinal})
turns into:

\begin{equation}
  -\frac{2\nabla^2\phi}{c^2}=\kappa' \left[\frac{8\pi
    GL_M^{2(1-b)}\rho}{c^2(b-2)}\right] ^{1/b}.
\label{exacta1}
\end{equation}

  Since the acceleration is defined by: $ | \boldsymbol{a}| := | \nabla\phi | $,
then

\begin{equation}
  -\frac{2}{c^2}\nabla \cdot\boldsymbol{a}=\kappa'\left[\frac{8\pi
    GL_M^{2(1-b)}\rho}{c^2(b-2)}\right] ^{1/b}.
\label{exacta}
\end{equation}

This last equation will allow us to fix the parameter $b$ such that it
is possible to recover a MONDian acceleration (\ref{mondian}).

\section{Recovering MOND}
\label{mond}

At order of magnitude, eq.~(\ref{exacta}) turns into:

\begin{equation}
  \frac{a}{c^2r}\approx\left[\frac{GL_M^{2-2b}\rho}{c^2}\right]^{1/b}. 
\label{aprox}
\end{equation}

  For a point mass source located at the origin, the density $\rho$ is given by:

\begin{equation}
  \rho = M \delta(\boldsymbol{r}),
\label{densidad}
\end{equation}

\noindent where $\delta(\boldsymbol{r})$ is the three-dimensional Dirac's
delta distribution in spherical coordinates.  Approximating the previous
equation to the same order of magnitude yields $\rho\approx M/r^3$,
and so expression~(\ref{aprox}) reduces to

\begin{equation}
  a\approx (GM)^{1/b}L_M^{2(1-b)/b}c^{2(b-1)/b}r^{(b-3)/b}. 
\label{aproximado}
\end{equation}

  On the one hand, the flattening of rotation curves requires $a \propto
r^{-1} $, and so:

\begin{equation}
  b=3/2.
\label{resultadob}
\end{equation}

\noindent On the other hand, the weakest field limit of approximation yields a
non-relativistic description of gravity and as such, the velocity of light
should not appear on eq.~\eqref{aproximado}.  In other words,

\begin{equation}
  L_M \propto c.
\label{demandaLM}
\end{equation}

  As noted by~\citep{mendozatula}, using dimensional analysis in the
description of a point mass source for a relativistic version of MOND,
it is possible to construct two independent fundamental lengths and it
is expected that  the length \( L_M \) should be a function of those two
lengths, in other words:

\begin{gather} L_M:= \zeta r_g^{\alpha} l_M^{\beta},
		\label{lmapprox} \\ \intertext{where} r_g :=
\frac{GM}{c^2}, \qquad l_M:=\left(\frac{GM}{a_0}\right)^{1/2},
		\label{longitudes} \end{gather}

\noindent represent the gravitational radius and a MONDian ``mass-length''
scale respectively.  The constant \( \zeta \) is a proportionality factor
and the exponents \( \alpha \) and \( \beta \) must satisfy the condition 
$\alpha+\beta=1$ so that eq.~\eqref{lmapprox}  is dimensionally
correct.  With the aid of eq.~\eqref{demandaLM} it follows that
$\alpha=-1/2$, and so, $\beta=3/2$. In other words:

\begin{equation}
  L_M=\zeta \left(\frac{GM}{a_0^3}\right)^{1/4}c.
\label{LMaprox}
\end{equation}

Using this expression for $L_M$ and the value for $b$ previously found, 
at order of magnitude the acceleration~\eqref{aproximado} reaches a 
MONDian value: \( a \approx ( GM a_0 )^{1/2}  / r \).

  In order to fully show that a MONDian non-relativistic limit
is obtained in the weak-field limit of the theory, we proceed as
follows. Direct substitution of the values obtained for $b$ and $L_M$,
into eq. (\ref{exacta}) yields:

\begin{equation}
  -2\nabla \cdot \boldsymbol{a} = \kappa'\left(a_0GM\right)^{1/2} \left[ \frac{
    4 \delta(r) }{ \zeta r^2 } \right]^{2/3}.
\label{valores}
\end{equation}

\noindent where we have used the fact that the three-dimensional 
Dirac's delta function is given by 
$\delta(\boldsymbol{r})=\delta(r)/4\pi r^2$. 

 For Schwartz distributions it is impossible in general terms, to define a
product in such a way that the resulting  distribution forms an algebra
with acceptable topological properties~\citep{Schwartz}. Schwartz's
impossibility result states that it is not possible to have a differential
algebra that contains the space of distributions and preserves the
product of continuous functions.  To overcome these disadvantages,
\citep{Colombeau,Colombeau2} has developed a theory of
generalised functions, which allows to define a fully consistent product
of distributions.  As such, we can consider   Dirac's delta distribution
as a standard function so that  we can write the following identity

\begin{equation}
  [\delta(r)]^{2/3} = [\delta(r)]^{-1/3} \ \delta(r).
\label{diracdelta}
\end{equation}

With this relation, eq. (\ref{valores}) turns into:

\begin{equation}
  -2\nabla \cdot\boldsymbol{a} = \kappa'
    \left( a_0 G M \right)^{1/2} \left( \frac{4}{\zeta} 
    \right)^{2/3}\left[\frac{1}{r^4\delta(r)} 
    \right]^{1/3}\delta(r).
\label{fder}
\end{equation}

  Since we are searching for a MONDian value for the acceleration, let us 
assume it obeys the following general power law:

\begin{equation}
  \boldsymbol{a}=\lambda r^\sigma \boldsymbol{e_r},
\label{formaa}
\end{equation}

\noindent where $\boldsymbol{e_r}$ is a unitary vector in the radial
direction, \( \lambda \) and \( \sigma \)  are constants so that:

\begin{equation}
  \nabla\cdot \boldsymbol{a} = \lambda(\sigma+2)r^{\sigma-1}.
\label{divergencia}
\end{equation}

Substitution of this last equation into~(\ref{fder}) and performing an
integration over $r$ yields:

\begin{equation}
-\frac{2\lambda(\sigma+2)}{\sigma}r^\sigma\bigg|_{r=0}^{r=\infty}=\kappa'\left(a_0GM\right)^{1/2}
	\left(\frac{4}{\zeta}\right)^{2/3}
	\left.\left[\frac{1}{r^4\delta(r)}\right]^{1/3}\right|_{r=0}.
	\label{integracion}
\end{equation}

  Let us now use the fact that $\delta(0)$ can be obtained from the
following relation~\citep{delta}:

\begin{equation}
  \delta(r=0) = \lim_{r\rightarrow 0}\frac{1}{2\pi r},
\label{delta0}
\end{equation}

\noindent and substitute it into eq.~(\ref{integracion}) in order to
obtain: 

\begin{equation}
  -\frac{ \lambda(\sigma+2)}{\gamma}r^\sigma \bigg|_{r=0}^{r=\infty} = 
    \kappa'\left(a_0GM\right)^{1/2}\left(\frac{4\pi}{\zeta^2}\right)^{1/3}
    \left.\frac{1}{r}\right|_{r=0}.
\label{compararr}
\end{equation}

\noindent Since $\zeta$ and $\lambda$ are constants, the following relation
is necessarily satisfied:

\begin{equation}
  \sigma = -1,
\end{equation}

\noindent which is an expected result from the order of magnitude analysis
developed above in order to obtain flat rotation curves.
Equation~(\ref{compararr}) is then reduced to:

\begin{equation}
  -\lambda=\kappa'\left(a_0GM\right)^{1/2}\left(\frac{4\pi}{
    \zeta^2}\right)^{1/3}.
\label{lambda}
\end{equation}

  In order to recover a MONDian acceleration~(\ref{mondian}) limit, it is 
necessary that $\lambda=-\left(a_0GM\right)^{1/2}$ and so:

\begin{equation}
	\zeta=2\left(\kappa'^3\pi\right)^{1/2}.
	\label{zeta}
\end{equation}

\section{Second order perturbation analysis}
\label{PPN}

  In order to show that a MONDian solution is directly obtained from
the field equations of the previous analysis, let us proceed as follows.
Substituting the value $b=3/2$ in eq.~\eqref{campo1} and~\eqref{traza1},
the field equations and the trace take the following form:

\begin{gather}
	3{\cal{R}}^{1/2} {\cal{R}}_{\mu\nu}-g_{\mu\nu}{\cal{R}}^{3/2}
		=\frac{16\pi G}{c^4 L_M}T_{\mu\nu},
	\label{campo1b}
	\\
	\intertext{and}
	-{\cal{R}}^{3/2}=\frac{16\pi G}{c^4 L_M}T.
	\label{traza1b}
\end{gather}

\noindent This last equation is meaningless unless the energy-momentum
tensor is defined with a minus sign on the right-hand side of
eq.~\eqref{variacionmateria}.  This fact is closely related to the multiple
branches that the solution space of any \( F(R) \) theory of gravity has,
which is usually ascribed to the choice of the Riemann tensor~(see e.g.
the discussion on the appendix of \cite{mendozalensing}).  Quite curiously
for the previous and following discussions it is not necessary at all to
enter into further discussions about this, since the obtained results 
require only the square of eq.~\eqref{traza1b}.
Substituting eqs.~\eqref{traza1b}, \eqref{LMaprox}~and~\eqref{Taylor1}
into \eqref{campo1b}, we obtain the following 
field equations:

\begin{equation}
	3(R_{\mu\nu}+H_{\mu\nu})=\left(\frac{16\pi }{c^5 \zeta}\right)^{2/3}
		\frac{(a_0G)^{1/2}}{M^{1/6}}\frac{\left(T g_{\mu\nu}-T_{\mu\nu}\right)}{T^{1/3}}.
	\label{componentes}
\end{equation}

  If we now perturb the metric \( g_{\mu\nu} \) about a flat Minkowsky
space-time \( \eta_{\mu\nu} \) we obtain:

\begin{equation}
	g_{\mu\nu}= \eta_{\mu\nu}+ \xi_{\mu\nu}.
	\label{ppnmetric}
\end{equation}

\noindent The quantity $\xi_{\mu\nu}$ is the perturbation
expanded in powers of $1/c$. To first order in $\xi$ (second order
in $1/c$), the time and space components of $R_{\mu\nu}$ are
\citep{Will}:

\begin{gather}
	{}^{(2)}R_{00}=\frac{1}{2} \nabla^2 \xi_{00},
	\label{temporal}
	\\
	\intertext{and}
	{}^{(2)}R_{ij}=\frac{1}{2} \nabla^2 \xi_{ij},
	\label{espacial}
\end{gather}

\noindent where we have suppressed the upper index in $\xi$ in the 
understanding that only the second perturbation order 
is relevant in our analysis. We have also chosen the 
PPN gauge for which: $\xi^\mu\,_{i,\mu}-1/2 \xi^\mu\,_{\mu,i}=0$.

The constraint eq.~\eqref{lineal} implies that $H_{\mu\nu}$ is a
linear function in ${\cal{R}}$ and so by eq.~\eqref{Rlineal} it is 
also linear $R$, but now the 
proportionality constant is a second rank tensor $\kappa_{\mu\nu}$, i.e.:

\begin{equation}
	H_{\mu\nu}= \kappa_{\mu\nu}{\cal{R}}.
	\label{linealHmunu}
\end{equation}

\noindent Thus, if the first perturbative term of $R$ is a second order
term, $H_{\mu\nu}$ would also be as such.

The spatial components of eq.~\eqref{componentes} are:

\begin{equation}
	3\left(\frac{1}{2}\nabla^2 \xi_{ij}+{}^{(2)}H_{ij}\right)=
		\left(\frac{16\pi }{c^5 \zeta}\right)^{2/3}
		\frac{(a_0G)^{1/2}}{M^{1/6}}\frac{\left(T g_{ij}-T_{ij}\right)}{T^{1/3}}.
		\label{espaciales}
\end{equation}

\noindent The left-hand side of this relation is of second order and so,
to obtain a second order term on the right-hand side, the last factor
involving only T must be of order $ \mathcal{O}(c^{4/3})$.  For dust,
the lowest perturbation order on \( T \) implies that: $T=\rho c^2$ and
$T_{ij}=0$, satisfying the previous requirement.  This is a consistency
check that our proposal is coherent at the lowest perturbation
order. Thus, for dust and a  point mass source, eq.~\eqref{espaciales}
turns into:

\begin{equation}
	3\left(\frac{1}{2}\nabla^2 \xi_{ij}+{}^{(2)}H_{ij}\right)=
		-\left(\frac{4\delta(r)}{r^2 \zeta}\right)^{2/3}
		\frac{(a_0GM)^{1/2}}{c^2}\delta_{ij}.
		\label{comparacion1}
\end{equation}

\noindent Comparison of this expression with \eqref{valores},
yields:

\begin{equation}
	3\left(\frac{1}{2}\nabla^2 \xi_{ij}+{}^{(2)}H_{ij}\right)=
		\frac{2\nabla^2 \phi}{\kappa' c^2}\delta_{ij}.
		\label{comparacionpotencial}
\end{equation}

\noindent In order to recover the value of \( \xi_{ij} \) consistent with
an isotropic metric, i.e.  $\xi_{ij}=2\phi/c^2\delta_{ij}$, 
the following value of $H_{ij}$ is obtained:

\begin{equation}
	{}^{(2)}H_{ij}=\frac{\nabla^2 \phi}{c^2}\delta_{ij}
		\left(\frac{2}{3\kappa'} - 1\right).
	\label{Hij}
\end{equation}

\noindent An analogous procedure for the time component yields:

\begin{equation}
	{}^{(2)}H_{00}=-\frac{\nabla^2 \phi}{c^2}.
	\label{H00}
\end{equation}

Physically in a weak field limit it is expected that the Jordan and
the Einstein frames, with metrics \( g_{\mu\nu} \) and \( h_{\mu\nu} \)
respectively, lead to the same physical results. This means that 
the contributions of the tensor $H_{\mu\nu}$ must be sufficiently small.
Bearing this in mind and the arbitrariness of the constant \( \kappa' \),
let us choose:

\begin{equation}
	\kappa'=\frac{2}{3}, \qquad \text{for which:} \qquad \kappa=\frac{1}{3}.
\label{kappa}
\end{equation}

\noindent Using these results together with eqs.~\eqref{linealHmunu},
\eqref{Hij}, \eqref{Rlineal} and~\eqref{H00}, we obtain that 
the only non vanish component of the tensor $\kappa_{\mu\nu}$ is 
$\kappa_{00}=1/3$.

Finally, from eq.~\eqref{zeta}, we find:

\begin{equation}
	\zeta=\left(\frac{32 \pi}{27}\right)^{1/2}.
\end{equation}

\section{Discussion}
\label{Conclusiones}

  It has become quite challenging to find a general expression that could
potentially yield MOND on the weak-field limit of
approximation~\citep{Babichev, scalar-tensor-sanders, Bekenstein-milgrom,
Bekenstein, Skordis, Zhao, Sanders-t-v-s, Bruneton, Zlosnik, Milgrom-bimetric,
Deffayet}.   Many of the proposal fail since the metric
coefficients~\eqref{segundoorden} at second perturbation order are in no 
agreement with the mathematical particularities of the theories involved.
Most importantly, it has always been desired that the field equations of
a relativistic version of MOND are of the second order and involve only a
power law function of the Ricci scalar.  In this article, we have shown how
to build such a second order field equations theory based on the metric
coefficients~\eqref{segundoorden} that converges to the simplest form of
MOND~\eqref{mondian} on its weakest limit of approximation 

  It is worth noticing at this point that the developed formalism in this
article is such that the ``coupling constant length'' \( L_M \) of the
gravitational action~\eqref{action} is a proportional to 
\( M^{1/4} \).
\citep{Sobouti,mendoza-guevara,mendozatula} have all encountered this
particularity when trying to build relativistic versions of MOND for metric
formulations of gravity.   Since it is customary that the gravitational
action does not depend on the mass (or the energy-momentum tensor) then
these authors have noticed that ``one should not be surprised if
some of the commonly accepted notions, even at the fundamental level of
the action, require generalisations and re-thinking''.  An extended metric
theory of gravity goes beyond the traditional general relativity ideas
and in this way, we should change some of our standard views regarding its
fundamental principles.  Accepting this we can formally write the
gravitational action \( S_g \) -first term on the right-hand side of
eq.~\eqref{action}- inspired by the generalisations made by~\citep{Harko1,
Harko2, Harko3, Lobo, Harko4} and
following a similar approach as that of~\citep{Carranza-torres}:

\begin{gather}
  S_g = - \frac{ c^3 }{ 16 \pi G } \int{ \frac{ f(\chi)  }{ L_M^2 } \,
    \sqrt{-g} \, \mathrm{d}x^4 },
    			\label{action-tipo-harko} \\
  \intertext{where following the results of eq.~\eqref{LMaprox}:}
  \begin{split}
    L_M &= c \left( \frac{ G }{ a_0^3 } \right)^{1/4} 
      \int{ \rho \mathrm{d}^3 x }, \\
        &= \frac{ 1 }{ c } \,  \left( \frac{ G }{ a_0^3
          } \right)^{1/4} \int{  {\cal{L}}_\text{matt} \mathrm{d}^3 x },
  \end{split}
			\label{lm-nuevo} 
\end{gather}

\noindent and we have used the fact that the matter Lagrangian \(
{\cal{L}}_\text{matt} = \rho c^2 \) for dust, and for systems with
sufficient degree of symmetry, e.g. isotropic or spherically symmetric
space-times, the integral is taken over all the causally connected
masses related to a particular problem.  For the single point mass
source discussed in this article, \( \rho = M \delta(\boldsymbol{r})
\) and in this case, eq.~\eqref{action-tipo-harko} converges to the
gravitational action~\eqref{action}.

  At this point, it is important to note that usually in the analysis
of $F(R)$ theories on a Post-Newtonian frame, the comparison
with a Brans-Dicke-like scalar-tensor theory can be achieved
\citep{OlmoPalatini}. In our work, we do not appeal to this analogy and
we keep the original equations throughout our analysis.

We choose to work in the frame of the Palatini formalism since it provides
a deeper understanding of our proposal than the metric formalism because we
do not restrict to a special kind of connection.  While it is true that in
standard general relativity, the Palatini formalism does not seem to bring something new, its use
in areas where general relativity is not tested has been 
extended~\citep{palatini1, palatini2, palatini3}.

  The value of the parameter $ b = 3/2 $ required for an extended metric
theory of gravity \(  f(\chi) = \chi^b \) in the Palatini formalism to
yield a MONDian behaviour has appeared on many other works related to the
cosmology \cite{SalvatoreQ1,SalvatoreQ2} and to MOND using a pure metric
approach to the problem \cite{mendozatula} using Noether's symmetry.
It is quite interesting that this value also appears in the Palatini
formalism presented in this article and together with the previous
findings sheds some light into a deepest understanding of gravitational
phenomena beyond Einstein's general relativity.

  The analysis performed in this article shows that it is possible to
explain the flattening of rotation curves and the Tully-fisher law
from our $f(\chi)=\chi^{3/2}$ theory using the  Palatini formalism.
By construction it not only reproduces the dynamics of material particles
required to flatten rotation curves which show a Tully-Fisher scaling,
but also reproduces bending of light associated to individual,
groups and clusters of galaxies.  This approach can be tested in
cosmological models dealing with the accelerated expansion of the
universe and in complex gravitational lensing, such as for example
the ones produced by collisional clusters of galaxies.  The fourth
perturbation order of the theory can be also used to model the dynamics
of clusters of galaxies in a completely analogous way as it was done
in \cite{sergio-tula-oliver}. These analyses are beyond the scope of
this article and will be studied elsewhere.

\section*{Acknowledgements} This work was supported by DGAPA-UNAM
 (IN112616)
and CONACyT (240512) grants. EB and SM acknowledge economic support from
CONACyT (517586 and 26344).

\bibliographystyle{apsrev4-1}
\bibliography{palatiniMOND}

\end{document}

%% file: aas-journals.tex
\newcommand{\aj}{Astronomical Journal}
\newcommand{\araa}{Ann. Rev. Ast. \& Ast.}
\newcommand{\apjl}{ApJL}

\newcommand{\aap}{Astronomy and Astrophysics}
\newcommand{\aapr}{Astronomy and Astrophysics Reviews}

\newcommand{\jcap}{Journal of Cosmology and Astroparticle Physics}

\newcommand{\mnras}{MNRAS}

\newcommand{\bain}{Bulletin Astronomical Institute of the Netherlands}

\newcommand{\physrep}{Physics Reports}

%% file: palatiniMOND.bbl
\begin{thebibliography}{89}%
\makeatletter
\providecommand \@ifxundefined [1]{%
 \@ifx{#1\undefined}
}%
\providecommand \@ifnum [1]{%
 \ifnum #1\expandafter \@firstoftwo
 \else \expandafter \@secondoftwo
 \fi
}%
\providecommand \@ifx [1]{%
 \ifx #1\expandafter \@firstoftwo
 \else \expandafter \@secondoftwo
 \fi
}%
\providecommand \natexlab [1]{#1}%
\providecommand \enquote  [1]{``#1''}%
\providecommand \bibnamefont  [1]{#1}%
\providecommand \bibfnamefont [1]{#1}%
\providecommand \citenamefont [1]{#1}%
\providecommand \href@noop [0]{\@secondoftwo}%
\providecommand \href [0]{\begingroup \@sanitize@url \@href}%
\providecommand \@href[1]{\@@startlink{#1}\@@href}%
\providecommand \@@href[1]{\endgroup#1\@@endlink}%
\providecommand \@sanitize@url [0]{\catcode `\\12\catcode `\$12\catcode
  `\&12\catcode `\#12\catcode `\^12\catcode `\_12\catcode `\%12\relax}%
\providecommand \@@startlink[1]{}%
\providecommand \@@endlink[0]{}%
\providecommand \url  [0]{\begingroup\@sanitize@url \@url }%
\providecommand \@url [1]{\endgroup\@href {#1}{\urlprefix }}%
\providecommand \urlprefix  [0]{URL }%
\providecommand \Eprint [0]{\href }%
\providecommand \doibase [0]{http://dx.doi.org/}%
\providecommand \selectlanguage [0]{\@gobble}%
\providecommand \bibinfo  [0]{\@secondoftwo}%
\providecommand \bibfield  [0]{\@secondoftwo}%
\providecommand \translation [1]{[#1]}%
\providecommand \BibitemOpen [0]{}%
\providecommand \bibitemStop [0]{}%
\providecommand \bibitemNoStop [0]{.\EOS\space}%
\providecommand \EOS [0]{\spacefactor3000\relax}%
\providecommand \BibitemShut  [1]{\csname bibitem#1\endcsname}%
\let\auto@bib@innerbib\@empty
\bibitem [{\citenamefont {{Bernal}}\ \emph {et~al.}(2011)\citenamefont
  {{Bernal}}, \citenamefont {{Capozziello}}, \citenamefont {{Hidalgo}},\ and\
  \citenamefont {{Mendoza}}}]{mendozatula}%
  \BibitemOpen
  \bibfield  {author} {\bibinfo {author} {\bibfnamefont {T.}~\bibnamefont
  {{Bernal}}}, \bibinfo {author} {\bibfnamefont {S.}~\bibnamefont
  {{Capozziello}}}, \bibinfo {author} {\bibfnamefont {J.~C.}\ \bibnamefont
  {{Hidalgo}}}, \ and\ \bibinfo {author} {\bibfnamefont {S.}~\bibnamefont
  {{Mendoza}}},\ }\href {\doibase 10.1140/epjc/s10052-011-1794-z} {\bibfield
  {journal} {\bibinfo  {journal} {European Physical Journal C}\ }\textbf
  {\bibinfo {volume} {71}},\ \bibinfo {eid} {1794} (\bibinfo {year} {2011})},\
  \Eprint {http://arxiv.org/abs/1108.5588} {arXiv:1108.5588 [astro-ph.CO]}
  \BibitemShut {NoStop}%
\bibitem [{\citenamefont {{Will}}(1971{\natexlab{a}})}]{Will1}%
  \BibitemOpen
  \bibfield  {author} {\bibinfo {author} {\bibfnamefont {C.~M.}\ \bibnamefont
  {{Will}}},\ }\href {\doibase 10.1086/150804} {\bibfield  {journal} {\bibinfo
  {journal} {\apj}\ }\textbf {\bibinfo {volume} {163}},\ \bibinfo {pages} {611}
  (\bibinfo {year} {1971}{\natexlab{a}})}\BibitemShut {NoStop}%
\bibitem [{\citenamefont {{Will}}(1971{\natexlab{b}})}]{Will2}%
  \BibitemOpen
  \bibfield  {author} {\bibinfo {author} {\bibfnamefont {C.~M.}\ \bibnamefont
  {{Will}}},\ }\href {\doibase 10.1086/151124} {\bibfield  {journal} {\bibinfo
  {journal} {\apj}\ }\textbf {\bibinfo {volume} {169}},\ \bibinfo {pages} {125}
  (\bibinfo {year} {1971}{\natexlab{b}})}\BibitemShut {NoStop}%
\bibitem [{\citenamefont {{Will}}(1971{\natexlab{c}})}]{Will3}%
  \BibitemOpen
  \bibfield  {author} {\bibinfo {author} {\bibfnamefont {C.~M.}\ \bibnamefont
  {{Will}}},\ }\href {\doibase 10.1086/150906} {\bibfield  {journal} {\bibinfo
  {journal} {\apj}\ }\textbf {\bibinfo {volume} {165}},\ \bibinfo {pages} {409}
  (\bibinfo {year} {1971}{\natexlab{c}})}\BibitemShut {NoStop}%
\bibitem [{\citenamefont {{Will}}(1971{\natexlab{d}})}]{Will4}%
  \BibitemOpen
  \bibfield  {author} {\bibinfo {author} {\bibfnamefont {C.~M.}\ \bibnamefont
  {{Will}}},\ }\href {\doibase 10.1086/151125} {\bibfield  {journal} {\bibinfo
  {journal} {\apj}\ }\textbf {\bibinfo {volume} {169}},\ \bibinfo {pages} {141}
  (\bibinfo {year} {1971}{\natexlab{d}})}\BibitemShut {NoStop}%
\bibitem [{\citenamefont {{Hulse}}\ and\ \citenamefont
  {{Taylor}}(1975)}]{TaylorHulse}%
  \BibitemOpen
  \bibfield  {author} {\bibinfo {author} {\bibfnamefont {R.~A.}\ \bibnamefont
  {{Hulse}}}\ and\ \bibinfo {author} {\bibfnamefont {J.~H.}\ \bibnamefont
  {{Taylor}}},\ }\href {\doibase 10.1086/181708} {\bibfield  {journal}
  {\bibinfo  {journal} {\apjl}\ }\textbf {\bibinfo {volume} {195}},\ \bibinfo
  {pages} {L51} (\bibinfo {year} {1975})}\BibitemShut {NoStop}%
\bibitem [{\citenamefont {{Kramer}}\ \emph {et~al.}(2006)\citenamefont
  {{Kramer}}, \citenamefont {{Stairs}}, \citenamefont {{Manchester}},
  \citenamefont {{McLaughlin}}, \citenamefont {{Lyne}}, \citenamefont
  {{Ferdman}}, \citenamefont {{Burgay}}, \citenamefont {{Lorimer}},
  \citenamefont {{Possenti}}, \citenamefont {{D'Amico}}, \citenamefont
  {{Sarkissian}}, \citenamefont {{Hobbs}}, \citenamefont {{Reynolds}},
  \citenamefont {{Freire}},\ and\ \citenamefont {{Camilo}}}]{Kramer1}%
  \BibitemOpen
  \bibfield  {author} {\bibinfo {author} {\bibfnamefont {M.}~\bibnamefont
  {{Kramer}}}, \bibinfo {author} {\bibfnamefont {I.~H.}\ \bibnamefont
  {{Stairs}}}, \bibinfo {author} {\bibfnamefont {R.~N.}\ \bibnamefont
  {{Manchester}}}, \bibinfo {author} {\bibfnamefont {M.~A.}\ \bibnamefont
  {{McLaughlin}}}, \bibinfo {author} {\bibfnamefont {A.~G.}\ \bibnamefont
  {{Lyne}}}, \bibinfo {author} {\bibfnamefont {R.~D.}\ \bibnamefont
  {{Ferdman}}}, \bibinfo {author} {\bibfnamefont {M.}~\bibnamefont {{Burgay}}},
  \bibinfo {author} {\bibfnamefont {D.~R.}\ \bibnamefont {{Lorimer}}}, \bibinfo
  {author} {\bibfnamefont {A.}~\bibnamefont {{Possenti}}}, \bibinfo {author}
  {\bibfnamefont {N.}~\bibnamefont {{D'Amico}}}, \bibinfo {author}
  {\bibfnamefont {J.~M.}\ \bibnamefont {{Sarkissian}}}, \bibinfo {author}
  {\bibfnamefont {G.~B.}\ \bibnamefont {{Hobbs}}}, \bibinfo {author}
  {\bibfnamefont {J.~E.}\ \bibnamefont {{Reynolds}}}, \bibinfo {author}
  {\bibfnamefont {P.~C.~C.}\ \bibnamefont {{Freire}}}, \ and\ \bibinfo {author}
  {\bibfnamefont {F.}~\bibnamefont {{Camilo}}},\ }\href {\doibase
  10.1126/science.1132305} {\bibfield  {journal} {\bibinfo  {journal}
  {Science}\ }\textbf {\bibinfo {volume} {314}},\ \bibinfo {pages} {97}
  (\bibinfo {year} {2006})},\ \Eprint {http://arxiv.org/abs/astro-ph/0609417}
  {astro-ph/0609417} \BibitemShut {NoStop}%
\bibitem [{\citenamefont {{Kramer}}\ and\ \citenamefont
  {{Champion}}(2013)}]{Kramer2}%
  \BibitemOpen
  \bibfield  {author} {\bibinfo {author} {\bibfnamefont {M.}~\bibnamefont
  {{Kramer}}}\ and\ \bibinfo {author} {\bibfnamefont {D.~J.}\ \bibnamefont
  {{Champion}}},\ }\href {\doibase 10.1088/0264-9381/30/22/224009} {\bibfield
  {journal} {\bibinfo  {journal} {Classical and Quantum Gravity}\ }\textbf
  {\bibinfo {volume} {30}},\ \bibinfo {eid} {224009} (\bibinfo {year}
  {2013})}\BibitemShut {NoStop}%
\bibitem [{\citenamefont {{Kramer}}\ \emph
  {et~al.}(2005{\natexlab{a}})\citenamefont {{Kramer}}, \citenamefont
  {{Lorimer}}, \citenamefont {{Lyne}}, \citenamefont {{McLaughlin}},
  \citenamefont {{Burgay}}, \citenamefont {{D'Amico}}, \citenamefont
  {{Possenti}}, \citenamefont {{Camilo}}, \citenamefont {{Freire}},
  \citenamefont {{Joshi}}, \citenamefont {{Manchester}}, \citenamefont
  {{Reynolds}}, \citenamefont {{Sarkissian Australia Telescope National
  Facility}}, \citenamefont {{Csiro}}, \citenamefont {{Stairs}},\ and\
  \citenamefont {{Ferdman}}}]{Kramer3}%
  \BibitemOpen
  \bibfield  {author} {\bibinfo {author} {\bibfnamefont {M.}~\bibnamefont
  {{Kramer}}}, \bibinfo {author} {\bibfnamefont {D.~R.}\ \bibnamefont
  {{Lorimer}}}, \bibinfo {author} {\bibfnamefont {A.~G.}\ \bibnamefont
  {{Lyne}}}, \bibinfo {author} {\bibfnamefont {M.}~\bibnamefont
  {{McLaughlin}}}, \bibinfo {author} {\bibfnamefont {M.}~\bibnamefont
  {{Burgay}}}, \bibinfo {author} {\bibfnamefont {N.}~\bibnamefont {{D'Amico}}},
  \bibinfo {author} {\bibfnamefont {A.}~\bibnamefont {{Possenti}}}, \bibinfo
  {author} {\bibfnamefont {F.}~\bibnamefont {{Camilo}}}, \bibinfo {author}
  {\bibfnamefont {P.~C.~C.}\ \bibnamefont {{Freire}}}, \bibinfo {author}
  {\bibfnamefont {B.~C.}\ \bibnamefont {{Joshi}}}, \bibinfo {author}
  {\bibfnamefont {R.~N.}\ \bibnamefont {{Manchester}}}, \bibinfo {author}
  {\bibfnamefont {J.}~\bibnamefont {{Reynolds}}}, \bibinfo {author}
  {\bibfnamefont {J.}~\bibnamefont {{Sarkissian Australia Telescope National
  Facility}}}, \bibinfo {author} {\bibnamefont {{Csiro}}}, \bibinfo {author}
  {\bibfnamefont {A.~I.~H.}\ \bibnamefont {{Stairs}}}, \ and\ \bibinfo {author}
  {\bibfnamefont {R.~D.}\ \bibnamefont {{Ferdman}}},\ }in\ \href@noop {} {\emph
  {\bibinfo {booktitle} {22nd Texas Symposium on Relativistic Astrophysics}}},\
  \bibinfo {editor} {edited by\ \bibinfo {editor} {\bibfnamefont
  {P.}~\bibnamefont {{Chen}}}, \bibinfo {editor} {\bibfnamefont
  {E.}~\bibnamefont {{Bloom}}}, \bibinfo {editor} {\bibfnamefont
  {G.}~\bibnamefont {{Madejski}}}, \ and\ \bibinfo {editor} {\bibfnamefont
  {V.}~\bibnamefont {{Patrosian}}}}\ (\bibinfo {year} {2005})\ pp.\ \bibinfo
  {pages} {142--148},\ \Eprint {http://arxiv.org/abs/astro-ph/0503386}
  {astro-ph/0503386} \BibitemShut {NoStop}%
\bibitem [{\citenamefont {{Kramer}}\ \emph
  {et~al.}(2005{\natexlab{b}})\citenamefont {{Kramer}}, \citenamefont {{Lyne}},
  \citenamefont {{Burgay}}, \citenamefont {{Possenti}}, \citenamefont
  {{Manchester}}, \citenamefont {{Camilo}}, \citenamefont {{McLaughlin}},
  \citenamefont {{Lorimer}}, \citenamefont {{D'Amico}}, \citenamefont
  {{Joshi}}, \citenamefont {{Reynolds}},\ and\ \citenamefont
  {{Freire}}}]{Kramer4}%
  \BibitemOpen
  \bibfield  {author} {\bibinfo {author} {\bibfnamefont {M.}~\bibnamefont
  {{Kramer}}}, \bibinfo {author} {\bibfnamefont {A.~G.}\ \bibnamefont
  {{Lyne}}}, \bibinfo {author} {\bibfnamefont {M.}~\bibnamefont {{Burgay}}},
  \bibinfo {author} {\bibfnamefont {A.}~\bibnamefont {{Possenti}}}, \bibinfo
  {author} {\bibfnamefont {R.~N.}\ \bibnamefont {{Manchester}}}, \bibinfo
  {author} {\bibfnamefont {F.}~\bibnamefont {{Camilo}}}, \bibinfo {author}
  {\bibfnamefont {M.~A.}\ \bibnamefont {{McLaughlin}}}, \bibinfo {author}
  {\bibfnamefont {D.~R.}\ \bibnamefont {{Lorimer}}}, \bibinfo {author}
  {\bibfnamefont {N.}~\bibnamefont {{D'Amico}}}, \bibinfo {author}
  {\bibfnamefont {B.~C.}\ \bibnamefont {{Joshi}}}, \bibinfo {author}
  {\bibfnamefont {J.}~\bibnamefont {{Reynolds}}}, \ and\ \bibinfo {author}
  {\bibfnamefont {P.~C.~C.}\ \bibnamefont {{Freire}}},\ }in\ \href@noop {}
  {\emph {\bibinfo {booktitle} {Binary Radio Pulsars}}},\ \bibinfo {series}
  {Astronomical Society of the Pacific Conference Series}, Vol.\ \bibinfo
  {volume} {328},\ \bibinfo {editor} {edited by\ \bibinfo {editor}
  {\bibfnamefont {F.~A.}\ \bibnamefont {{Rasio}}}\ and\ \bibinfo {editor}
  {\bibfnamefont {I.~H.}\ \bibnamefont {{Stairs}}}}\ (\bibinfo {year} {2005})\
  p.~\bibinfo {pages} {59},\ \Eprint {http://arxiv.org/abs/astro-ph/0405179}
  {astro-ph/0405179} \BibitemShut {NoStop}%
\bibitem [{\citenamefont {{Kramer}}(2013)}]{Kramer5}%
  \BibitemOpen
  \bibfield  {author} {\bibinfo {author} {\bibfnamefont {M.}~\bibnamefont
  {{Kramer}}},\ }in\ \href {\doibase 10.1017/S174392131202306X} {\emph
  {\bibinfo {booktitle} {Neutron Stars and Pulsars: Challenges and
  Opportunities after 80 years}}},\ \bibinfo {series} {IAU Symposium}, Vol.\
  \bibinfo {volume} {291},\ \bibinfo {editor} {edited by\ \bibinfo {editor}
  {\bibfnamefont {J.}~\bibnamefont {{van Leeuwen}}}}\ (\bibinfo {year} {2013})\
  pp.\ \bibinfo {pages} {19--26},\ \Eprint {http://arxiv.org/abs/1211.2457}
  {arXiv:1211.2457 [astro-ph.HE]} \BibitemShut {NoStop}%
\bibitem [{\citenamefont {{Oort}}(1932)}]{Oort}%
  \BibitemOpen
  \bibfield  {author} {\bibinfo {author} {\bibfnamefont {J.~H.}\ \bibnamefont
  {{Oort}}},\ }\href@noop {} {\bibfield  {journal} {\bibinfo  {journal}
  {\bain}\ }\textbf {\bibinfo {volume} {6}},\ \bibinfo {pages} {249} (\bibinfo
  {year} {1932})}\BibitemShut {NoStop}%
\bibitem [{\citenamefont {{Zwicky}}(1933)}]{Zwicky}%
  \BibitemOpen
  \bibfield  {author} {\bibinfo {author} {\bibfnamefont {F.}~\bibnamefont
  {{Zwicky}}},\ }\href@noop {} {\bibfield  {journal} {\bibinfo  {journal}
  {Helvetica Physica Acta}\ }\textbf {\bibinfo {volume} {6}},\ \bibinfo {pages}
  {110} (\bibinfo {year} {1933})}\BibitemShut {NoStop}%
\bibitem [{\citenamefont {{Ostriker}}\ and\ \citenamefont
  {{Peebles}}(1973)}]{Peebles}%
  \BibitemOpen
  \bibfield  {author} {\bibinfo {author} {\bibfnamefont {J.~P.}\ \bibnamefont
  {{Ostriker}}}\ and\ \bibinfo {author} {\bibfnamefont {P.~J.~E.}\ \bibnamefont
  {{Peebles}}},\ }\href {\doibase 10.1086/152513} {\bibfield  {journal}
  {\bibinfo  {journal} {\apj}\ }\textbf {\bibinfo {volume} {186}},\ \bibinfo
  {pages} {467} (\bibinfo {year} {1973})}\BibitemShut {NoStop}%
\bibitem [{\citenamefont {{Bosma}}(1981)}]{Bosma}%
  \BibitemOpen
  \bibfield  {author} {\bibinfo {author} {\bibfnamefont {A.}~\bibnamefont
  {{Bosma}}},\ }\href {\doibase 10.1086/113063} {\bibfield  {journal} {\bibinfo
   {journal} {\aj}\ }\textbf {\bibinfo {volume} {86}},\ \bibinfo {pages} {1825}
  (\bibinfo {year} {1981})}\BibitemShut {NoStop}%
\bibitem [{\citenamefont {{Rubin}}\ \emph {et~al.}(1982)\citenamefont
  {{Rubin}}, \citenamefont {{Ford}}, \citenamefont {{Thonnard}},\ and\
  \citenamefont {{Burstein}}}]{Rubin}%
  \BibitemOpen
  \bibfield  {author} {\bibinfo {author} {\bibfnamefont {V.~C.}\ \bibnamefont
  {{Rubin}}}, \bibinfo {author} {\bibfnamefont {W.~K.}\ \bibnamefont {{Ford}},
  \bibfnamefont {Jr.}}, \bibinfo {author} {\bibfnamefont {N.}~\bibnamefont
  {{Thonnard}}}, \ and\ \bibinfo {author} {\bibfnamefont {D.}~\bibnamefont
  {{Burstein}}},\ }\href {\doibase 10.1086/160355} {\bibfield  {journal}
  {\bibinfo  {journal} {\apj}\ }\textbf {\bibinfo {volume} {261}},\ \bibinfo
  {pages} {439} (\bibinfo {year} {1982})}\BibitemShut {NoStop}%
\bibitem [{\citenamefont {{Gunn}}\ \emph {et~al.}(1978)\citenamefont {{Gunn}},
  \citenamefont {{Lee}}, \citenamefont {{Lerche}}, \citenamefont {{Schramm}},\
  and\ \citenamefont {{Steigman}}}]{Gunn}%
  \BibitemOpen
  \bibfield  {author} {\bibinfo {author} {\bibfnamefont {J.~E.}\ \bibnamefont
  {{Gunn}}}, \bibinfo {author} {\bibfnamefont {B.~W.}\ \bibnamefont {{Lee}}},
  \bibinfo {author} {\bibfnamefont {I.}~\bibnamefont {{Lerche}}}, \bibinfo
  {author} {\bibfnamefont {D.~N.}\ \bibnamefont {{Schramm}}}, \ and\ \bibinfo
  {author} {\bibfnamefont {G.}~\bibnamefont {{Steigman}}},\ }\href {\doibase
  10.1086/156335} {\bibfield  {journal} {\bibinfo  {journal} {\apj}\ }\textbf
  {\bibinfo {volume} {223}},\ \bibinfo {pages} {1015} (\bibinfo {year}
  {1978})}\BibitemShut {NoStop}%
\bibitem [{\citenamefont {{Vittorio}}\ and\ \citenamefont
  {{Silk}}(1984)}]{Vittorio}%
  \BibitemOpen
  \bibfield  {author} {\bibinfo {author} {\bibfnamefont {N.}~\bibnamefont
  {{Vittorio}}}\ and\ \bibinfo {author} {\bibfnamefont {J.}~\bibnamefont
  {{Silk}}},\ }\href {\doibase 10.1086/184361} {\bibfield  {journal} {\bibinfo
  {journal} {\apjl}\ }\textbf {\bibinfo {volume} {285}},\ \bibinfo {pages}
  {L39} (\bibinfo {year} {1984})}\BibitemShut {NoStop}%
\bibitem [{\citenamefont {{Tremaine}}\ and\ \citenamefont
  {{Gunn}}(1979)}]{Gunn-tremaine}%
  \BibitemOpen
  \bibfield  {author} {\bibinfo {author} {\bibfnamefont {S.}~\bibnamefont
  {{Tremaine}}}\ and\ \bibinfo {author} {\bibfnamefont {J.~E.}\ \bibnamefont
  {{Gunn}}},\ }\href {\doibase 10.1103/PhysRevLett.42.407} {\bibfield
  {journal} {\bibinfo  {journal} {Physical Review Letters}\ }\textbf {\bibinfo
  {volume} {42}},\ \bibinfo {pages} {407} (\bibinfo {year} {1979})}\BibitemShut
  {NoStop}%
\bibitem [{\citenamefont {{Bond}}\ and\ \citenamefont
  {{Efstathiou}}(1984)}]{Bond-efs}%
  \BibitemOpen
  \bibfield  {author} {\bibinfo {author} {\bibfnamefont {J.~R.}\ \bibnamefont
  {{Bond}}}\ and\ \bibinfo {author} {\bibfnamefont {G.}~\bibnamefont
  {{Efstathiou}}},\ }\href {\doibase 10.1086/184362} {\bibfield  {journal}
  {\bibinfo  {journal} {\apjl}\ }\textbf {\bibinfo {volume} {285}},\ \bibinfo
  {pages} {L45} (\bibinfo {year} {1984})}\BibitemShut {NoStop}%
\bibitem [{\citenamefont {{Blumenthal}}\ \emph {et~al.}(1984)\citenamefont
  {{Blumenthal}}, \citenamefont {{Faber}}, \citenamefont {{Primack}},\ and\
  \citenamefont {{Rees}}}]{Blumenthal}%
  \BibitemOpen
  \bibfield  {author} {\bibinfo {author} {\bibfnamefont {G.~R.}\ \bibnamefont
  {{Blumenthal}}}, \bibinfo {author} {\bibfnamefont {S.~M.}\ \bibnamefont
  {{Faber}}}, \bibinfo {author} {\bibfnamefont {J.~R.}\ \bibnamefont
  {{Primack}}}, \ and\ \bibinfo {author} {\bibfnamefont {M.~J.}\ \bibnamefont
  {{Rees}}},\ }\href {\doibase 10.1038/311517a0} {\bibfield  {journal}
  {\bibinfo  {journal} {\nat}\ }\textbf {\bibinfo {volume} {311}},\ \bibinfo
  {pages} {517} (\bibinfo {year} {1984})}\BibitemShut {NoStop}%
\bibitem [{\citenamefont {{White}}\ \emph {et~al.}(1983)\citenamefont
  {{White}}, \citenamefont {{Frenk}},\ and\ \citenamefont
  {{Davis}}}]{Neutrinos}%
  \BibitemOpen
  \bibfield  {author} {\bibinfo {author} {\bibfnamefont {S.~D.~M.}\
  \bibnamefont {{White}}}, \bibinfo {author} {\bibfnamefont {C.~S.}\
  \bibnamefont {{Frenk}}}, \ and\ \bibinfo {author} {\bibfnamefont
  {M.}~\bibnamefont {{Davis}}},\ }\href {\doibase 10.1086/184139} {\bibfield
  {journal} {\bibinfo  {journal} {\apjl}\ }\textbf {\bibinfo {volume} {274}},\
  \bibinfo {pages} {L1} (\bibinfo {year} {1983})}\BibitemShut {NoStop}%
\bibitem [{\citenamefont {{Frenk}}\ \emph {et~al.}(1988)\citenamefont
  {{Frenk}}, \citenamefont {{White}}, \citenamefont {{Davis}},\ and\
  \citenamefont {{Efstathiou}}}]{Frenk-white}%
  \BibitemOpen
  \bibfield  {author} {\bibinfo {author} {\bibfnamefont {C.~S.}\ \bibnamefont
  {{Frenk}}}, \bibinfo {author} {\bibfnamefont {S.~D.~M.}\ \bibnamefont
  {{White}}}, \bibinfo {author} {\bibfnamefont {M.}~\bibnamefont {{Davis}}}, \
  and\ \bibinfo {author} {\bibfnamefont {G.}~\bibnamefont {{Efstathiou}}},\
  }\href {\doibase 10.1086/166213} {\bibfield  {journal} {\bibinfo  {journal}
  {\apj}\ }\textbf {\bibinfo {volume} {327}},\ \bibinfo {pages} {507} (\bibinfo
  {year} {1988})}\BibitemShut {NoStop}%
\bibitem [{\citenamefont {{Milgrom}}(1983{\natexlab{a}})}]{Milgrom1}%
  \BibitemOpen
  \bibfield  {author} {\bibinfo {author} {\bibfnamefont {M.}~\bibnamefont
  {{Milgrom}}},\ }\href {\doibase 10.1086/161130} {\bibfield  {journal}
  {\bibinfo  {journal} {\apj}\ }\textbf {\bibinfo {volume} {270}},\ \bibinfo
  {pages} {365} (\bibinfo {year} {1983}{\natexlab{a}})}\BibitemShut {NoStop}%
\bibitem [{\citenamefont {{Milgrom}}(1983{\natexlab{b}})}]{Milgrom2}%
  \BibitemOpen
  \bibfield  {author} {\bibinfo {author} {\bibfnamefont {M.}~\bibnamefont
  {{Milgrom}}},\ }\href {\doibase 10.1086/161131} {\bibfield  {journal}
  {\bibinfo  {journal} {\apj}\ }\textbf {\bibinfo {volume} {270}},\ \bibinfo
  {pages} {371} (\bibinfo {year} {1983}{\natexlab{b}})}\BibitemShut {NoStop}%
\bibitem [{\citenamefont {{Sotiriou}}\ and\ \citenamefont
  {{Faraoni}}(2010)}]{f(R)}%
  \BibitemOpen
  \bibfield  {author} {\bibinfo {author} {\bibfnamefont {T.~P.}\ \bibnamefont
  {{Sotiriou}}}\ and\ \bibinfo {author} {\bibfnamefont {V.}~\bibnamefont
  {{Faraoni}}},\ }\href {\doibase 10.1103/RevModPhys.82.451} {\bibfield
  {journal} {\bibinfo  {journal} {Reviews of Modern Physics}\ }\textbf
  {\bibinfo {volume} {82}},\ \bibinfo {pages} {451} (\bibinfo {year} {2010})},\
  \Eprint {http://arxiv.org/abs/0805.1726} {arXiv:0805.1726 [gr-qc]}
  \BibitemShut {NoStop}%
\bibitem [{\citenamefont {{Sotiriou}}\ and\ \citenamefont
  {{Liberati}}(2007)}]{liberati}%
  \BibitemOpen
  \bibfield  {author} {\bibinfo {author} {\bibfnamefont {T.~P.}\ \bibnamefont
  {{Sotiriou}}}\ and\ \bibinfo {author} {\bibfnamefont {S.}~\bibnamefont
  {{Liberati}}},\ }\href {\doibase 10.1016/j.aop.2006.06.002} {\bibfield
  {journal} {\bibinfo  {journal} {Annals of Physics}\ }\textbf {\bibinfo
  {volume} {322}},\ \bibinfo {pages} {935} (\bibinfo {year} {2007})},\ \Eprint
  {http://arxiv.org/abs/gr-qc/0604006} {gr-qc/0604006} \BibitemShut {NoStop}%
\bibitem [{\citenamefont {{Capozziello}}\ and\ \citenamefont
  {{Faraoni}}(2011)}]{capozziello}%
  \BibitemOpen
  \bibfield  {author} {\bibinfo {author} {\bibfnamefont {S.}~\bibnamefont
  {{Capozziello}}}\ and\ \bibinfo {author} {\bibfnamefont {V.}~\bibnamefont
  {{Faraoni}}},\ }\href {\doibase 10.1007/978-94-007-0165-6} {\emph {\bibinfo
  {title} {Beyond Einstein Gravity: A Survey of Gravitational Theories for
  Cosmology and Astrophysics, Fundamental Theories of Physics, Volume 170.~ISBN
  978-94-007-0164-9.~Springer Science+Business Media B.V., 2011}}}\ (\bibinfo
  {year} {2011})\BibitemShut {NoStop}%
\bibitem [{\citenamefont {{Capozziello}}\ and\ \citenamefont {{de
  Laurentis}}(2011)}]{Salvatoreextendido}%
  \BibitemOpen
  \bibfield  {author} {\bibinfo {author} {\bibfnamefont {S.}~\bibnamefont
  {{Capozziello}}}\ and\ \bibinfo {author} {\bibfnamefont {M.}~\bibnamefont
  {{de Laurentis}}},\ }\href {\doibase 10.1016/j.physrep.2011.09.003}
  {\bibfield  {journal} {\bibinfo  {journal} {\physrep}\ }\textbf {\bibinfo
  {volume} {509}},\ \bibinfo {pages} {167} (\bibinfo {year} {2011})},\ \Eprint
  {http://arxiv.org/abs/1108.6266} {arXiv:1108.6266 [gr-qc]} \BibitemShut
  {NoStop}%
\bibitem [{\citenamefont {{Nojiri}}\ and\ \citenamefont
  {{Odintsov}}(2011)}]{nojiri11}%
  \BibitemOpen
  \bibfield  {author} {\bibinfo {author} {\bibfnamefont {S.}~\bibnamefont
  {{Nojiri}}}\ and\ \bibinfo {author} {\bibfnamefont {S.~D.}\ \bibnamefont
  {{Odintsov}}},\ }\href {\doibase 10.1016/j.physrep.2011.04.001} {\bibfield
  {journal} {\bibinfo  {journal} {\physrep}\ }\textbf {\bibinfo {volume}
  {505}},\ \bibinfo {pages} {59} (\bibinfo {year} {2011})},\ \Eprint
  {http://arxiv.org/abs/1011.0544} {arXiv:1011.0544 [gr-qc]} \BibitemShut
  {NoStop}%
\bibitem [{\citenamefont {{Bekenstein}}(2004)}]{Bekenstein}%
  \BibitemOpen
  \bibfield  {author} {\bibinfo {author} {\bibfnamefont {J.~D.}\ \bibnamefont
  {{Bekenstein}}},\ }\href {\doibase 10.1103/PhysRevD.70.083509} {\bibfield
  {journal} {\bibinfo  {journal} {\prd}\ }\textbf {\bibinfo {volume} {70}},\
  \bibinfo {eid} {083509} (\bibinfo {year} {2004})},\ \Eprint
  {http://arxiv.org/abs/astro-ph/0403694} {astro-ph/0403694} \BibitemShut
  {NoStop}%
\bibitem [{\citenamefont {{Bekenstein}}\ and\ \citenamefont
  {{Sanders}}(2012)}]{Bek2}%
  \BibitemOpen
  \bibfield  {author} {\bibinfo {author} {\bibfnamefont {J.~D.}\ \bibnamefont
  {{Bekenstein}}}\ and\ \bibinfo {author} {\bibfnamefont {R.~H.}\ \bibnamefont
  {{Sanders}}},\ }\href {\doibase 10.1111/j.1745-3933.2011.01206.x} {\bibfield
  {journal} {\bibinfo  {journal} {\mnras}\ }\textbf {\bibinfo {volume} {421}},\
  \bibinfo {pages} {L59} (\bibinfo {year} {2012})},\ \Eprint
  {http://arxiv.org/abs/1110.5048} {arXiv:1110.5048 [astro-ph.CO]} \BibitemShut
  {NoStop}%
\bibitem [{\citenamefont {{Sanders}}(2005)}]{Sanders-t-v-s}%
  \BibitemOpen
  \bibfield  {author} {\bibinfo {author} {\bibfnamefont {R.~H.}\ \bibnamefont
  {{Sanders}}},\ }\href {\doibase 10.1111/j.1365-2966.2005.09375.x} {\bibfield
  {journal} {\bibinfo  {journal} {\mnras}\ }\textbf {\bibinfo {volume} {363}},\
  \bibinfo {pages} {459} (\bibinfo {year} {2005})},\ \Eprint
  {http://arxiv.org/abs/astro-ph/0502222} {astro-ph/0502222} \BibitemShut
  {NoStop}%
\bibitem [{\citenamefont {{Zlosnik}}\ \emph {et~al.}(2007)\citenamefont
  {{Zlosnik}}, \citenamefont {{Ferreira}},\ and\ \citenamefont
  {{Starkman}}}]{Zlosnik}%
  \BibitemOpen
  \bibfield  {author} {\bibinfo {author} {\bibfnamefont {T.~G.}\ \bibnamefont
  {{Zlosnik}}}, \bibinfo {author} {\bibfnamefont {P.~G.}\ \bibnamefont
  {{Ferreira}}}, \ and\ \bibinfo {author} {\bibfnamefont {G.~D.}\ \bibnamefont
  {{Starkman}}},\ }\href {\doibase 10.1103/PhysRevD.75.044017} {\bibfield
  {journal} {\bibinfo  {journal} {\prd}\ }\textbf {\bibinfo {volume} {75}},\
  \bibinfo {eid} {044017} (\bibinfo {year} {2007})},\ \Eprint
  {http://arxiv.org/abs/astro-ph/0607411} {astro-ph/0607411} \BibitemShut
  {NoStop}%
\bibitem [{\citenamefont {{Skordis}}(2009)}]{SkortisTeVeS}%
  \BibitemOpen
  \bibfield  {author} {\bibinfo {author} {\bibfnamefont {C.}~\bibnamefont
  {{Skordis}}},\ }\href {\doibase 10.1088/0264-9381/26/14/143001} {\bibfield
  {journal} {\bibinfo  {journal} {Classical and Quantum Gravity}\ }\textbf
  {\bibinfo {volume} {26}},\ \bibinfo {eid} {143001} (\bibinfo {year}
  {2009})},\ \Eprint {http://arxiv.org/abs/0903.3602} {arXiv:0903.3602
  [astro-ph.CO]} \BibitemShut {NoStop}%
\bibitem [{\citenamefont {{Babichev}}\ \emph {et~al.}(2011)\citenamefont
  {{Babichev}}, \citenamefont {{Deffayet}},\ and\ \citenamefont
  {{Esposito-Far{\`e}se}}}]{Babichev}%
  \BibitemOpen
  \bibfield  {author} {\bibinfo {author} {\bibfnamefont {E.}~\bibnamefont
  {{Babichev}}}, \bibinfo {author} {\bibfnamefont {C.}~\bibnamefont
  {{Deffayet}}}, \ and\ \bibinfo {author} {\bibfnamefont {G.}~\bibnamefont
  {{Esposito-Far{\`e}se}}},\ }\href {\doibase 10.1103/PhysRevD.84.061502}
  {\bibfield  {journal} {\bibinfo  {journal} {\prd}\ }\textbf {\bibinfo
  {volume} {84}},\ \bibinfo {eid} {061502} (\bibinfo {year} {2011})},\ \Eprint
  {http://arxiv.org/abs/1106.2538} {arXiv:1106.2538 [gr-qc]} \BibitemShut
  {NoStop}%
\bibitem [{\citenamefont {{Milgrom}}(2009)}]{Milgrom-bimetric}%
  \BibitemOpen
  \bibfield  {author} {\bibinfo {author} {\bibfnamefont {M.}~\bibnamefont
  {{Milgrom}}},\ }\href {\doibase 10.1103/PhysRevD.80.123536} {\bibfield
  {journal} {\bibinfo  {journal} {\prd}\ }\textbf {\bibinfo {volume} {80}},\
  \bibinfo {eid} {123536} (\bibinfo {year} {2009})},\ \Eprint
  {http://arxiv.org/abs/0912.0790} {arXiv:0912.0790 [gr-qc]} \BibitemShut
  {NoStop}%
\bibitem [{\citenamefont {{Clifton}}\ and\ \citenamefont
  {{Zlosnik}}(2010)}]{Bimetric2}%
  \BibitemOpen
  \bibfield  {author} {\bibinfo {author} {\bibfnamefont {T.}~\bibnamefont
  {{Clifton}}}\ and\ \bibinfo {author} {\bibfnamefont {T.~G.}\ \bibnamefont
  {{Zlosnik}}},\ }\href {\doibase 10.1103/PhysRevD.81.103525} {\bibfield
  {journal} {\bibinfo  {journal} {\prd}\ }\textbf {\bibinfo {volume} {81}},\
  \bibinfo {eid} {103525} (\bibinfo {year} {2010})},\ \Eprint
  {http://arxiv.org/abs/1002.1448} {arXiv:1002.1448} \BibitemShut {NoStop}%
\bibitem [{\citenamefont {Demir}\ and\ \citenamefont
  {Karahan}(2014)}]{demir2014}%
  \BibitemOpen
  \bibfield  {author} {\bibinfo {author} {\bibfnamefont {D.~A.}\ \bibnamefont
  {Demir}}\ and\ \bibinfo {author} {\bibfnamefont {C.~N.}\ \bibnamefont
  {Karahan}},\ }\href {\doibase 10.1140/epjc/s10052-014-3204-9} {\bibfield
  {journal} {\bibinfo  {journal} {The European Physical Journal C}\ }\textbf
  {\bibinfo {volume} {74}},\ \bibinfo {pages} {1} (\bibinfo {year}
  {2014})}\BibitemShut {NoStop}%
\bibitem [{\citenamefont {{Blanchet}}(2007{\natexlab{a}})}]{Blanchet1}%
  \BibitemOpen
  \bibfield  {author} {\bibinfo {author} {\bibfnamefont {L.}~\bibnamefont
  {{Blanchet}}},\ }\href {\doibase 10.1088/0264-9381/24/14/002} {\bibfield
  {journal} {\bibinfo  {journal} {Classical and Quantum Gravity}\ }\textbf
  {\bibinfo {volume} {24}},\ \bibinfo {pages} {3541} (\bibinfo {year}
  {2007}{\natexlab{a}})},\ \Eprint {http://arxiv.org/abs/gr-qc/0609121}
  {gr-qc/0609121} \BibitemShut {NoStop}%
\bibitem [{\citenamefont {{Blanchet}}(2007{\natexlab{b}})}]{Blanchet2}%
  \BibitemOpen
  \bibfield  {author} {\bibinfo {author} {\bibfnamefont {L.}~\bibnamefont
  {{Blanchet}}},\ }\href {\doibase 10.1088/0264-9381/24/14/001} {\bibfield
  {journal} {\bibinfo  {journal} {Classical and Quantum Gravity}\ }\textbf
  {\bibinfo {volume} {24}},\ \bibinfo {pages} {3529} (\bibinfo {year}
  {2007}{\natexlab{b}})},\ \Eprint {http://arxiv.org/abs/astro-ph/0605637}
  {astro-ph/0605637} \BibitemShut {NoStop}%
\bibitem [{\citenamefont {{Blanchet}}\ and\ \citenamefont {{Le
  Tiec}}(2008)}]{Blanchet3}%
  \BibitemOpen
  \bibfield  {author} {\bibinfo {author} {\bibfnamefont {L.}~\bibnamefont
  {{Blanchet}}}\ and\ \bibinfo {author} {\bibfnamefont {A.}~\bibnamefont {{Le
  Tiec}}},\ }\href {\doibase 10.1103/PhysRevD.78.024031} {\bibfield  {journal}
  {\bibinfo  {journal} {\prd}\ }\textbf {\bibinfo {volume} {78}},\ \bibinfo
  {eid} {024031} (\bibinfo {year} {2008})},\ \Eprint
  {http://arxiv.org/abs/0804.3518} {arXiv:0804.3518} \BibitemShut {NoStop}%
\bibitem [{\citenamefont {{Soussa}}\ and\ \citenamefont
  {{Woodard}}(2003)}]{Soussa}%
  \BibitemOpen
  \bibfield  {author} {\bibinfo {author} {\bibfnamefont {M.~E.}\ \bibnamefont
  {{Soussa}}}\ and\ \bibinfo {author} {\bibfnamefont {R.~P.}\ \bibnamefont
  {{Woodard}}},\ }\href {\doibase 10.1088/0264-9381/20/13/321} {\bibfield
  {journal} {\bibinfo  {journal} {Classical and Quantum Gravity}\ }\textbf
  {\bibinfo {volume} {20}},\ \bibinfo {pages} {2737} (\bibinfo {year}
  {2003})},\ \Eprint {http://arxiv.org/abs/astro-ph/0302030} {astro-ph/0302030}
  \BibitemShut {NoStop}%
\bibitem [{\citenamefont {{Deffayet}}\ \emph
  {et~al.}(2011{\natexlab{a}})\citenamefont {{Deffayet}}, \citenamefont
  {{Esposito-Far{\`e}se}},\ and\ \citenamefont {{Woodard}}}]{Esposito}%
  \BibitemOpen
  \bibfield  {author} {\bibinfo {author} {\bibfnamefont {C.}~\bibnamefont
  {{Deffayet}}}, \bibinfo {author} {\bibfnamefont {G.}~\bibnamefont
  {{Esposito-Far{\`e}se}}}, \ and\ \bibinfo {author} {\bibfnamefont {R.~P.}\
  \bibnamefont {{Woodard}}},\ }\href {\doibase 10.1103/PhysRevD.84.124054}
  {\bibfield  {journal} {\bibinfo  {journal} {\prd}\ }\textbf {\bibinfo
  {volume} {84}},\ \bibinfo {eid} {124054} (\bibinfo {year}
  {2011}{\natexlab{a}})},\ \Eprint {http://arxiv.org/abs/1106.4984}
  {arXiv:1106.4984 [gr-qc]} \BibitemShut {NoStop}%
\bibitem [{\citenamefont {{Famaey}}\ and\ \citenamefont
  {{McGaugh}}(2012)}]{famey}%
  \BibitemOpen
  \bibfield  {author} {\bibinfo {author} {\bibfnamefont {B.}~\bibnamefont
  {{Famaey}}}\ and\ \bibinfo {author} {\bibfnamefont {S.~S.}\ \bibnamefont
  {{McGaugh}}},\ }\href {\doibase 10.12942/lrr-2012-10} {\bibfield  {journal}
  {\bibinfo  {journal} {Living Reviews in Relativity}\ }\textbf {\bibinfo
  {volume} {15}},\ \bibinfo {pages} {10} (\bibinfo {year} {2012})},\ \Eprint
  {http://arxiv.org/abs/1112.3960} {arXiv:1112.3960} \BibitemShut {NoStop}%
\bibitem [{\citenamefont {{Hernandez}}\ \emph {et~al.}(2010)\citenamefont
  {{Hernandez}}, \citenamefont {{Mendoza}}, \citenamefont {{Suarez}},\ and\
  \citenamefont {{Bernal}}}]{Xavier1}%
  \BibitemOpen
  \bibfield  {author} {\bibinfo {author} {\bibfnamefont {X.}~\bibnamefont
  {{Hernandez}}}, \bibinfo {author} {\bibfnamefont {S.}~\bibnamefont
  {{Mendoza}}}, \bibinfo {author} {\bibfnamefont {T.}~\bibnamefont {{Suarez}}},
  \ and\ \bibinfo {author} {\bibfnamefont {T.}~\bibnamefont {{Bernal}}},\
  }\href {\doibase 10.1051/0004-6361/200913301} {\bibfield  {journal} {\bibinfo
   {journal} {\aap}\ }\textbf {\bibinfo {volume} {514}},\ \bibinfo {eid} {A101}
  (\bibinfo {year} {2010})},\ \Eprint {http://arxiv.org/abs/0904.1434}
  {arXiv:0904.1434 [astro-ph.GA]} \BibitemShut {NoStop}%
\bibitem [{\citenamefont {{Hernandez}}\ and\ \citenamefont
  {{Jim{\'e}nez}}(2012)}]{Xavier2}%
  \BibitemOpen
  \bibfield  {author} {\bibinfo {author} {\bibfnamefont {X.}~\bibnamefont
  {{Hernandez}}}\ and\ \bibinfo {author} {\bibfnamefont {M.~A.}\ \bibnamefont
  {{Jim{\'e}nez}}},\ }\href {\doibase 10.1088/0004-637X/750/1/9} {\bibfield
  {journal} {\bibinfo  {journal} {\apj}\ }\textbf {\bibinfo {volume} {750}},\
  \bibinfo {eid} {9} (\bibinfo {year} {2012})},\ \Eprint
  {http://arxiv.org/abs/1108.4021} {arXiv:1108.4021} \BibitemShut {NoStop}%
\bibitem [{\citenamefont {{Hernandez}}\ \emph {et~al.}(2013)\citenamefont
  {{Hernandez}}, \citenamefont {{Jim{\'e}nez}},\ and\ \citenamefont
  {{Allen}}}]{Xavier3}%
  \BibitemOpen
  \bibfield  {author} {\bibinfo {author} {\bibfnamefont {X.}~\bibnamefont
  {{Hernandez}}}, \bibinfo {author} {\bibfnamefont {M.~A.}\ \bibnamefont
  {{Jim{\'e}nez}}}, \ and\ \bibinfo {author} {\bibfnamefont {C.}~\bibnamefont
  {{Allen}}},\ }\href {\doibase 10.1093/mnras/sts263} {\bibfield  {journal}
  {\bibinfo  {journal} {\mnras}\ }\textbf {\bibinfo {volume} {428}},\ \bibinfo
  {pages} {3196} (\bibinfo {year} {2013})},\ \Eprint
  {http://arxiv.org/abs/1206.5024} {arXiv:1206.5024 [astro-ph.GA]} \BibitemShut
  {NoStop}%
\bibitem [{\citenamefont {{Hernandez}}\ \emph {et~al.}(2012)\citenamefont
  {{Hernandez}}, \citenamefont {{Jim{\'e}nez}},\ and\ \citenamefont
  {{Allen}}}]{Xavier4}%
  \BibitemOpen
  \bibfield  {author} {\bibinfo {author} {\bibfnamefont {X.}~\bibnamefont
  {{Hernandez}}}, \bibinfo {author} {\bibfnamefont {M.~A.}\ \bibnamefont
  {{Jim{\'e}nez}}}, \ and\ \bibinfo {author} {\bibfnamefont {C.}~\bibnamefont
  {{Allen}}},\ }\href {\doibase 10.1140/epjc/s10052-012-1884-6} {\bibfield
  {journal} {\bibinfo  {journal} {European Physical Journal C}\ }\textbf
  {\bibinfo {volume} {72}},\ \bibinfo {eid} {1884} (\bibinfo {year} {2012})},\
  \Eprint {http://arxiv.org/abs/1105.1873} {arXiv:1105.1873 [astro-ph.GA]}
  \BibitemShut {NoStop}%
\bibitem [{\citenamefont {{Mendoza}}(2012)}]{mendozareview}%
  \BibitemOpen
  \bibfield  {author} {\bibinfo {author} {\bibfnamefont {S.}~\bibnamefont
  {{Mendoza}}},\ }\href@noop {} {\bibfield  {journal} {\bibinfo  {journal}
  {ArXiv e-prints}\ } (\bibinfo {year} {2012})},\ \Eprint
  {http://arxiv.org/abs/1208.3408} {arXiv:1208.3408 [astro-ph.CO]} \BibitemShut
  {NoStop}%
\bibitem [{\citenamefont {{Mendoza}}\ and\ \citenamefont
  {{Olmo}}(2014)}]{sergioyolmo}%
  \BibitemOpen
  \bibfield  {author} {\bibinfo {author} {\bibfnamefont {S.}~\bibnamefont
  {{Mendoza}}}\ and\ \bibinfo {author} {\bibfnamefont {G.~J.}\ \bibnamefont
  {{Olmo}}},\ }\href@noop {} {\bibfield  {journal} {\bibinfo  {journal} {ArXiv
  e-prints}\ } (\bibinfo {year} {2014})},\ \Eprint
  {http://arxiv.org/abs/1401.5104} {arXiv:1401.5104 [gr-qc]} \BibitemShut
  {NoStop}%
\bibitem [{\citenamefont {{Sanders}}(1990)}]{Sanders}%
  \BibitemOpen
  \bibfield  {author} {\bibinfo {author} {\bibfnamefont {R.~H.}\ \bibnamefont
  {{Sanders}}},\ }\href {\doibase 10.1007/BF00873540} {\bibfield  {journal}
  {\bibinfo  {journal} {\aapr}\ }\textbf {\bibinfo {volume} {2}},\ \bibinfo
  {pages} {1} (\bibinfo {year} {1990})}\BibitemShut {NoStop}%
\bibitem [{\citenamefont {{McGaugh}}(2004)}]{Mcgaugh}%
  \BibitemOpen
  \bibfield  {author} {\bibinfo {author} {\bibfnamefont {S.~S.}\ \bibnamefont
  {{McGaugh}}},\ }\href {\doibase 10.1086/421338} {\bibfield  {journal}
  {\bibinfo  {journal} {\apj}\ }\textbf {\bibinfo {volume} {609}},\ \bibinfo
  {pages} {652} (\bibinfo {year} {2004})},\ \Eprint
  {http://arxiv.org/abs/astro-ph/0403610} {astro-ph/0403610} \BibitemShut
  {NoStop}%
\bibitem [{\citenamefont {{McGaugh}}(1999)}]{Mcgaugh2}%
  \BibitemOpen
  \bibfield  {author} {\bibinfo {author} {\bibfnamefont {S.}~\bibnamefont
  {{McGaugh}}},\ }in\ \href@noop {} {\emph {\bibinfo {booktitle} {Galaxy
  Dynamics - A Rutgers Symposium}}},\ \bibinfo {series} {Astronomical Society
  of the Pacific Conference Series}, Vol.\ \bibinfo {volume} {182},\ \bibinfo
  {editor} {edited by\ \bibinfo {editor} {\bibfnamefont {D.~R.}\ \bibnamefont
  {{Merritt}}}, \bibinfo {editor} {\bibfnamefont {M.}~\bibnamefont
  {{Valluri}}}, \ and\ \bibinfo {editor} {\bibfnamefont {J.~A.}\ \bibnamefont
  {{Sellwood}}}}\ (\bibinfo {year} {1999})\ \Eprint
  {http://arxiv.org/abs/astro-ph/9812327} {astro-ph/9812327} \BibitemShut
  {NoStop}%
\bibitem [{\citenamefont {{Scarpa}}(2003)}]{Scarpa}%
  \BibitemOpen
  \bibfield  {author} {\bibinfo {author} {\bibfnamefont {R.}~\bibnamefont
  {{Scarpa}}},\ }\href@noop {} {\bibfield  {journal} {\bibinfo  {journal}
  {ArXiv Astrophysics e-prints}\ } (\bibinfo {year} {2003})},\ \Eprint
  {http://arxiv.org/abs/astro-ph/0302445} {astro-ph/0302445} \BibitemShut
  {NoStop}%
\bibitem [{\citenamefont {{McGaugh}}(2005)}]{Mcgaugh3}%
  \BibitemOpen
  \bibfield  {author} {\bibinfo {author} {\bibfnamefont {S.~S.}\ \bibnamefont
  {{McGaugh}}},\ }\href {\doibase 10.1086/432968} {\bibfield  {journal}
  {\bibinfo  {journal} {\apj}\ }\textbf {\bibinfo {volume} {632}},\ \bibinfo
  {pages} {859} (\bibinfo {year} {2005})},\ \Eprint
  {http://arxiv.org/abs/astro-ph/0506750} {astro-ph/0506750} \BibitemShut
  {NoStop}%
\bibitem [{\citenamefont {{Sanders}}\ and\ \citenamefont
  {{McGaugh}}(2002)}]{Sanders2}%
  \BibitemOpen
  \bibfield  {author} {\bibinfo {author} {\bibfnamefont {R.~H.}\ \bibnamefont
  {{Sanders}}}\ and\ \bibinfo {author} {\bibfnamefont {S.~S.}\ \bibnamefont
  {{McGaugh}}},\ }\href {\doibase 10.1146/annurev.astro.40.060401.093923}
  {\bibfield  {journal} {\bibinfo  {journal} {\araa}\ }\textbf {\bibinfo
  {volume} {40}},\ \bibinfo {pages} {263} (\bibinfo {year} {2002})},\ \Eprint
  {http://arxiv.org/abs/astro-ph/0204521} {astro-ph/0204521} \BibitemShut
  {NoStop}%
\bibitem [{\citenamefont {{Milgrom}}(1989)}]{Milgrom3}%
  \BibitemOpen
  \bibfield  {author} {\bibinfo {author} {\bibfnamefont {M.}~\bibnamefont
  {{Milgrom}}},\ }\href {\doibase 10.1086/167184} {\bibfield  {journal}
  {\bibinfo  {journal} {\apj}\ }\textbf {\bibinfo {volume} {338}},\ \bibinfo
  {pages} {121} (\bibinfo {year} {1989})}\BibitemShut {NoStop}%
\bibitem [{\citenamefont {{Brada}}\ and\ \citenamefont
  {{Milgrom}}(1999)}]{Brada-Milgrom}%
  \BibitemOpen
  \bibfield  {author} {\bibinfo {author} {\bibfnamefont {R.}~\bibnamefont
  {{Brada}}}\ and\ \bibinfo {author} {\bibfnamefont {M.}~\bibnamefont
  {{Milgrom}}},\ }\href {\doibase 10.1086/307402} {\bibfield  {journal}
  {\bibinfo  {journal} {\apj}\ }\textbf {\bibinfo {volume} {519}},\ \bibinfo
  {pages} {590} (\bibinfo {year} {1999})},\ \Eprint
  {http://arxiv.org/abs/astro-ph/9811013} {astro-ph/9811013} \BibitemShut
  {NoStop}%
\bibitem [{\citenamefont {{Mendoza}}\ \emph {et~al.}(2013)\citenamefont
  {{Mendoza}}, \citenamefont {{Bernal}}, \citenamefont {{Hernandez}},
  \citenamefont {{Hidalgo}},\ and\ \citenamefont {{Torres}}}]{mendozalensing}%
  \BibitemOpen
  \bibfield  {author} {\bibinfo {author} {\bibfnamefont {S.}~\bibnamefont
  {{Mendoza}}}, \bibinfo {author} {\bibfnamefont {T.}~\bibnamefont {{Bernal}}},
  \bibinfo {author} {\bibfnamefont {X.}~\bibnamefont {{Hernandez}}}, \bibinfo
  {author} {\bibfnamefont {J.~C.}\ \bibnamefont {{Hidalgo}}}, \ and\ \bibinfo
  {author} {\bibfnamefont {L.~A.}\ \bibnamefont {{Torres}}},\ }\href {\doibase
  10.1093/mnras/stt752} {\bibfield  {journal} {\bibinfo  {journal} {\mnras}\
  }\textbf {\bibinfo {volume} {433}},\ \bibinfo {pages} {1802} (\bibinfo {year}
  {2013})},\ \Eprint {http://arxiv.org/abs/1208.6241} {arXiv:1208.6241
  [astro-ph.CO]} \BibitemShut {NoStop}%
\bibitem [{\citenamefont {{Bernal}}\ \emph {et~al.}(2015)\citenamefont
  {{Bernal}}, \citenamefont {{L{\'o}pez-Corona}},\ and\ \citenamefont
  {{Mendoza}}}]{sergio-tula-oliver}%
  \BibitemOpen
  \bibfield  {author} {\bibinfo {author} {\bibfnamefont {T.}~\bibnamefont
  {{Bernal}}}, \bibinfo {author} {\bibfnamefont {O.}~\bibnamefont
  {{L{\'o}pez-Corona}}}, \ and\ \bibinfo {author} {\bibfnamefont
  {S.}~\bibnamefont {{Mendoza}}},\ }\href@noop {} {\bibfield  {journal}
  {\bibinfo  {journal} {ArXiv e-prints}\ } (\bibinfo {year} {2015})},\ \Eprint
  {http://arxiv.org/abs/1505.00037} {arXiv:1505.00037} \BibitemShut {NoStop}%
\bibitem [{\citenamefont {{Carranza}}\ and\ \citenamefont
  {{Mendoza}}(2015)}]{mendoza-diego}%
  \BibitemOpen
  \bibfield  {author} {\bibinfo {author} {\bibfnamefont {D.~A.}\ \bibnamefont
  {{Carranza}}}\ and\ \bibinfo {author} {\bibfnamefont {S.}~\bibnamefont
  {{Mendoza}}},\ }\href {\doibase 10.4236/jmp.2015.66084} {\bibfield  {journal}
  {\bibinfo  {journal} {Journal of Modern Physics}\ }\textbf {\bibinfo {volume}
  {6}},\ \bibinfo {pages} {786} (\bibinfo {year} {2015})},\ \Eprint
  {http://arxiv.org/abs/1402.6026} {arXiv:1402.6026 [gr-qc]} \BibitemShut
  {NoStop}%
\bibitem [{\citenamefont {{Misner}}\ \emph {et~al.}(1973)\citenamefont
  {{Misner}}, \citenamefont {{Thorne}},\ and\ \citenamefont
  {{Wheeler}}}]{gravitation}%
  \BibitemOpen
  \bibfield  {author} {\bibinfo {author} {\bibfnamefont {C.~W.}\ \bibnamefont
  {{Misner}}}, \bibinfo {author} {\bibfnamefont {K.~S.}\ \bibnamefont
  {{Thorne}}}, \ and\ \bibinfo {author} {\bibfnamefont {J.~A.}\ \bibnamefont
  {{Wheeler}}},\ }\href@noop {} {\emph {\bibinfo {title} {San Francisco:
  W.H.~Freeman and Co., 1973}}}\ (\bibinfo {year} {1973})\BibitemShut {NoStop}%
\bibitem [{\citenamefont {{Carroll}}(2004)}]{carroll}%
  \BibitemOpen
  \bibfield  {author} {\bibinfo {author} {\bibfnamefont {S.~M.}\ \bibnamefont
  {{Carroll}}},\ }\href@noop {} {\emph {\bibinfo {title} {Spacetime and
  geometry / Sean Carroll.~San Francisco, CA, USA: Addison Wesley, ISBN
  0-8053-8732-3, 2004, XIV + 513 pp.}}}\ (\bibinfo {year} {2004})\BibitemShut
  {NoStop}%
\bibitem [{\citenamefont {{Will}}(1993)}]{Will}%
  \BibitemOpen
  \bibfield  {author} {\bibinfo {author} {\bibfnamefont {C.~M.}\ \bibnamefont
  {{Will}}},\ }\href@noop {} {\emph {\bibinfo {title} {Theory and Experiment in
  Gravitational Physics, by Clifford M.~Will, pp.~396.~ISBN
  0521439736.~Cambridge, UK: Cambridge University Press, March 1993.}}}\
  (\bibinfo {year} {1993})\BibitemShut {NoStop}%
\bibitem [{\citenamefont {Schwartz}(1954)}]{Schwartz}%
  \BibitemOpen
  \bibfield  {author} {\bibinfo {author} {\bibfnamefont {L.}~\bibnamefont
  {Schwartz}},\ }\href@noop {} {\bibfield  {journal} {\bibinfo  {journal} {C.
  R. Acad. Sci. Paris}\ }\textbf {\bibinfo {volume} {239}},\ \bibinfo {pages}
  {847} (\bibinfo {year} {1954})}\BibitemShut {NoStop}%
\bibitem [{\citenamefont {Colombeau}(1990)}]{Colombeau}%
  \BibitemOpen
  \bibfield  {author} {\bibinfo {author} {\bibfnamefont {J.-F.}\ \bibnamefont
  {Colombeau}},\ }\href {\doibase 10.1090/S0273-0979-1990-15919-1} {\bibfield
  {journal} {\bibinfo  {journal} {Bull. Amer. Math. Soc. (N.S.)}\ }\textbf
  {\bibinfo {volume} {23}},\ \bibinfo {pages} {251} (\bibinfo {year}
  {1990})}\BibitemShut {NoStop}%
\bibitem [{\citenamefont {Colombeau}(1985)}]{Colombeau2}%
  \BibitemOpen
  \bibfield  {author} {\bibinfo {author} {\bibfnamefont {J.-F.}\ \bibnamefont
  {Colombeau}},\ }\href@noop {} {\emph {\bibinfo {title} {Elementary
  introduction to new generalized functions}}},\ \bibinfo {series}
  {North-Holland Mathematics Studies}, Vol.\ \bibinfo {volume} {113}\ (\bibinfo
   {publisher} {North-Holland Publishing Co., Amsterdam},\ \bibinfo {year}
  {1985})\ pp.\ \bibinfo {pages} {xiii+281},\ \bibinfo {note} {notes on Pure
  Mathematics, 103}\BibitemShut {NoStop}%
\bibitem [{\citenamefont {{Gsponer}}(2008)}]{delta}%
  \BibitemOpen
  \bibfield  {author} {\bibinfo {author} {\bibfnamefont {A.}~\bibnamefont
  {{Gsponer}}},\ }\href@noop {} {\bibfield  {journal} {\bibinfo  {journal}
  {ArXiv e-prints}\ } (\bibinfo {year} {2008})},\ \Eprint
  {http://arxiv.org/abs/0809.2576} {arXiv:0809.2576 [math-ph]} \BibitemShut
  {NoStop}%
\bibitem [{\citenamefont {{Sanders}}(1997)}]{scalar-tensor-sanders}%
  \BibitemOpen
  \bibfield  {author} {\bibinfo {author} {\bibfnamefont {R.~H.}\ \bibnamefont
  {{Sanders}}},\ }\href@noop {} {\bibfield  {journal} {\bibinfo  {journal}
  {\apj}\ }\textbf {\bibinfo {volume} {480}},\ \bibinfo {pages} {492} (\bibinfo
  {year} {1997})},\ \Eprint {http://arxiv.org/abs/astro-ph/9612099}
  {astro-ph/9612099} \BibitemShut {NoStop}%
\bibitem [{\citenamefont {{Bekenstein}}\ and\ \citenamefont
  {{Milgrom}}(1984)}]{Bekenstein-milgrom}%
  \BibitemOpen
  \bibfield  {author} {\bibinfo {author} {\bibfnamefont {J.}~\bibnamefont
  {{Bekenstein}}}\ and\ \bibinfo {author} {\bibfnamefont {M.}~\bibnamefont
  {{Milgrom}}},\ }\href {\doibase 10.1086/162570} {\bibfield  {journal}
  {\bibinfo  {journal} {\apj}\ }\textbf {\bibinfo {volume} {286}},\ \bibinfo
  {pages} {7} (\bibinfo {year} {1984})}\BibitemShut {NoStop}%
\bibitem [{\citenamefont {{Skordis}}(2008)}]{Skordis}%
  \BibitemOpen
  \bibfield  {author} {\bibinfo {author} {\bibfnamefont {C.}~\bibnamefont
  {{Skordis}}},\ }\href {\doibase 10.1103/PhysRevD.77.123502} {\bibfield
  {journal} {\bibinfo  {journal} {\prd}\ }\textbf {\bibinfo {volume} {77}},\
  \bibinfo {eid} {123502} (\bibinfo {year} {2008})},\ \Eprint
  {http://arxiv.org/abs/0801.1985} {arXiv:0801.1985} \BibitemShut {NoStop}%
\bibitem [{\citenamefont {{Zhao}}(2007)}]{Zhao}%
  \BibitemOpen
  \bibfield  {author} {\bibinfo {author} {\bibfnamefont {H.}~\bibnamefont
  {{Zhao}}},\ }\href {\doibase 10.1142/S0218271807011759} {\bibfield  {journal}
  {\bibinfo  {journal} {International Journal of Modern Physics D}\ }\textbf
  {\bibinfo {volume} {16}},\ \bibinfo {pages} {2055} (\bibinfo {year}
  {2007})},\ \Eprint {http://arxiv.org/abs/astro-ph/0610056} {astro-ph/0610056}
  \BibitemShut {NoStop}%
\bibitem [{\citenamefont {{Bruneton}}\ and\ \citenamefont
  {{Esposito-Far{\`e}se}}(2007)}]{Bruneton}%
  \BibitemOpen
  \bibfield  {author} {\bibinfo {author} {\bibfnamefont {J.-P.}\ \bibnamefont
  {{Bruneton}}}\ and\ \bibinfo {author} {\bibfnamefont {G.}~\bibnamefont
  {{Esposito-Far{\`e}se}}},\ }\href {\doibase 10.1103/PhysRevD.76.124012}
  {\bibfield  {journal} {\bibinfo  {journal} {\prd}\ }\textbf {\bibinfo
  {volume} {76}},\ \bibinfo {eid} {124012} (\bibinfo {year}
  {2007})}\BibitemShut {NoStop}%
\bibitem [{\citenamefont {{Deffayet}}\ \emph
  {et~al.}(2011{\natexlab{b}})\citenamefont {{Deffayet}}, \citenamefont
  {{Esposito-Far{\`e}se}},\ and\ \citenamefont {{Woodard}}}]{Deffayet}%
  \BibitemOpen
  \bibfield  {author} {\bibinfo {author} {\bibfnamefont {C.}~\bibnamefont
  {{Deffayet}}}, \bibinfo {author} {\bibfnamefont {G.}~\bibnamefont
  {{Esposito-Far{\`e}se}}}, \ and\ \bibinfo {author} {\bibfnamefont {R.~P.}\
  \bibnamefont {{Woodard}}},\ }\href {\doibase 10.1103/PhysRevD.84.124054}
  {\bibfield  {journal} {\bibinfo  {journal} {\prd}\ }\textbf {\bibinfo
  {volume} {84}},\ \bibinfo {eid} {124054} (\bibinfo {year}
  {2011}{\natexlab{b}})},\ \Eprint {http://arxiv.org/abs/1106.4984}
  {arXiv:1106.4984 [gr-qc]} \BibitemShut {NoStop}%
\bibitem [{\citenamefont {{Sobouti}}(2007)}]{Sobouti}%
  \BibitemOpen
  \bibfield  {author} {\bibinfo {author} {\bibfnamefont {Y.}~\bibnamefont
  {{Sobouti}}},\ }\href {\doibase 10.1051/0004-6361:20065188} {\bibfield
  {journal} {\bibinfo  {journal} {\aap}\ }\textbf {\bibinfo {volume} {464}},\
  \bibinfo {pages} {921} (\bibinfo {year} {2007})},\ \Eprint
  {http://arxiv.org/abs/astro-ph/0603302} {astro-ph/0603302} \BibitemShut
  {NoStop}%
\bibitem [{\citenamefont {{Mendoza}}\ and\ \citenamefont
  {{Rosas-Guevara}}(2007)}]{mendoza-guevara}%
  \BibitemOpen
  \bibfield  {author} {\bibinfo {author} {\bibfnamefont {S.}~\bibnamefont
  {{Mendoza}}}\ and\ \bibinfo {author} {\bibfnamefont {Y.~M.}\ \bibnamefont
  {{Rosas-Guevara}}},\ }\href {\doibase 10.1051/0004-6361:20066787} {\bibfield
  {journal} {\bibinfo  {journal} {\aap}\ }\textbf {\bibinfo {volume} {472}},\
  \bibinfo {pages} {367} (\bibinfo {year} {2007})},\ \Eprint
  {http://arxiv.org/abs/astro-ph/0610390} {astro-ph/0610390} \BibitemShut
  {NoStop}%
\bibitem [{\citenamefont {{Harko}}\ \emph {et~al.}(2011)\citenamefont
  {{Harko}}, \citenamefont {{Lobo}}, \citenamefont {{Nojiri}},\ and\
  \citenamefont {{Odintsov}}}]{Harko1}%
  \BibitemOpen
  \bibfield  {author} {\bibinfo {author} {\bibfnamefont {T.}~\bibnamefont
  {{Harko}}}, \bibinfo {author} {\bibfnamefont {F.~S.~N.}\ \bibnamefont
  {{Lobo}}}, \bibinfo {author} {\bibfnamefont {S.}~\bibnamefont {{Nojiri}}}, \
  and\ \bibinfo {author} {\bibfnamefont {S.~D.}\ \bibnamefont {{Odintsov}}},\
  }\href {\doibase 10.1103/PhysRevD.84.024020} {\bibfield  {journal} {\bibinfo
  {journal} {\prd}\ }\textbf {\bibinfo {volume} {84}},\ \bibinfo {eid} {024020}
  (\bibinfo {year} {2011})},\ \Eprint {http://arxiv.org/abs/1104.2669}
  {arXiv:1104.2669 [gr-qc]} \BibitemShut {NoStop}%
\bibitem [{\citenamefont {{Haghani}}\ \emph {et~al.}(2013)\citenamefont
  {{Haghani}}, \citenamefont {{Harko}}, \citenamefont {{Lobo}}, \citenamefont
  {{Sepangi}},\ and\ \citenamefont {{Shahidi}}}]{Harko2}%
  \BibitemOpen
  \bibfield  {author} {\bibinfo {author} {\bibfnamefont {Z.}~\bibnamefont
  {{Haghani}}}, \bibinfo {author} {\bibfnamefont {T.}~\bibnamefont {{Harko}}},
  \bibinfo {author} {\bibfnamefont {F.~S.~N.}\ \bibnamefont {{Lobo}}}, \bibinfo
  {author} {\bibfnamefont {H.~R.}\ \bibnamefont {{Sepangi}}}, \ and\ \bibinfo
  {author} {\bibfnamefont {S.}~\bibnamefont {{Shahidi}}},\ }\href {\doibase
  10.1103/PhysRevD.88.044023} {\bibfield  {journal} {\bibinfo  {journal}
  {\prd}\ }\textbf {\bibinfo {volume} {88}},\ \bibinfo {eid} {044023} (\bibinfo
  {year} {2013})},\ \Eprint {http://arxiv.org/abs/1304.5957} {arXiv:1304.5957
  [gr-qc]} \BibitemShut {NoStop}%
\bibitem [{\citenamefont {{Harko}}\ \emph {et~al.}(2013)\citenamefont
  {{Harko}}, \citenamefont {{Lobo}},\ and\ \citenamefont
  {{Minazzoli}}}]{Harko3}%
  \BibitemOpen
  \bibfield  {author} {\bibinfo {author} {\bibfnamefont {T.}~\bibnamefont
  {{Harko}}}, \bibinfo {author} {\bibfnamefont {F.~S.~N.}\ \bibnamefont
  {{Lobo}}}, \ and\ \bibinfo {author} {\bibfnamefont {O.}~\bibnamefont
  {{Minazzoli}}},\ }\href {\doibase 10.1103/PhysRevD.87.047501} {\bibfield
  {journal} {\bibinfo  {journal} {\prd}\ }\textbf {\bibinfo {volume} {87}},\
  \bibinfo {eid} {047501} (\bibinfo {year} {2013})},\ \Eprint
  {http://arxiv.org/abs/1210.4218} {arXiv:1210.4218 [gr-qc]} \BibitemShut
  {NoStop}%
\bibitem [{\citenamefont {{Lobo}}\ and\ \citenamefont {{Harko}}(2012)}]{Lobo}%
  \BibitemOpen
  \bibfield  {author} {\bibinfo {author} {\bibfnamefont {F.~S.~N.}\
  \bibnamefont {{Lobo}}}\ and\ \bibinfo {author} {\bibfnamefont
  {T.}~\bibnamefont {{Harko}}},\ }\href@noop {} {\bibfield  {journal} {\bibinfo
   {journal} {ArXiv e-prints}\ } (\bibinfo {year} {2012})},\ \Eprint
  {http://arxiv.org/abs/1211.0426} {arXiv:1211.0426 [gr-qc]} \BibitemShut
  {NoStop}%
\bibitem [{\citenamefont {{Harko}}\ \emph {et~al.}(2014)\citenamefont
  {{Harko}}, \citenamefont {{Lobo}}, \citenamefont {{Otalora}},\ and\
  \citenamefont {{Saridakis}}}]{Harko4}%
  \BibitemOpen
  \bibfield  {author} {\bibinfo {author} {\bibfnamefont {T.}~\bibnamefont
  {{Harko}}}, \bibinfo {author} {\bibfnamefont {F.~S.~N.}\ \bibnamefont
  {{Lobo}}}, \bibinfo {author} {\bibfnamefont {G.}~\bibnamefont {{Otalora}}}, \
  and\ \bibinfo {author} {\bibfnamefont {E.~N.}\ \bibnamefont {{Saridakis}}},\
  }\href {\doibase 10.1103/PhysRevD.89.124036} {\bibfield  {journal} {\bibinfo
  {journal} {\prd}\ }\textbf {\bibinfo {volume} {89}},\ \bibinfo {eid} {124036}
  (\bibinfo {year} {2014})},\ \Eprint {http://arxiv.org/abs/1404.6212}
  {arXiv:1404.6212 [gr-qc]} \BibitemShut {NoStop}%
\bibitem [{\citenamefont {{Carranza}}\ \emph {et~al.}(2013)\citenamefont
  {{Carranza}}, \citenamefont {{Mendoza}},\ and\ \citenamefont
  {{Torres}}}]{Carranza-torres}%
  \BibitemOpen
  \bibfield  {author} {\bibinfo {author} {\bibfnamefont {D.~A.}\ \bibnamefont
  {{Carranza}}}, \bibinfo {author} {\bibfnamefont {S.}~\bibnamefont
  {{Mendoza}}}, \ and\ \bibinfo {author} {\bibfnamefont {L.~A.}\ \bibnamefont
  {{Torres}}},\ }\href {\doibase 10.1140/epjc/s10052-013-2282-4} {\bibfield
  {journal} {\bibinfo  {journal} {European Physical Journal C}\ }\textbf
  {\bibinfo {volume} {73}},\ \bibinfo {eid} {2282} (\bibinfo {year} {2013})},\
  \Eprint {http://arxiv.org/abs/1208.2502} {arXiv:1208.2502} \BibitemShut
  {NoStop}%
\bibitem [{\citenamefont {{Olmo}}(2005)}]{OlmoPalatini}%
  \BibitemOpen
  \bibfield  {author} {\bibinfo {author} {\bibfnamefont {G.~J.}\ \bibnamefont
  {{Olmo}}},\ }\href {\doibase 10.1103/PhysRevD.72.083505} {\bibfield
  {journal} {\bibinfo  {journal} {\prd}\ }\textbf {\bibinfo {volume} {72}},\
  \bibinfo {eid} {083505} (\bibinfo {year} {2005})},\ \Eprint
  {http://arxiv.org/abs/gr-qc/0505135} {gr-qc/0505135} \BibitemShut {NoStop}%
\bibitem [{\citenamefont {{Olmo}}\ \emph {et~al.}(2015)\citenamefont {{Olmo}},
  \citenamefont {{Rubiera-Garcia}},\ and\ \citenamefont
  {{Sanchez-Puente}}}]{palatini1}%
  \BibitemOpen
  \bibfield  {author} {\bibinfo {author} {\bibfnamefont {G.~J.}\ \bibnamefont
  {{Olmo}}}, \bibinfo {author} {\bibfnamefont {D.}~\bibnamefont
  {{Rubiera-Garcia}}}, \ and\ \bibinfo {author} {\bibfnamefont
  {A.}~\bibnamefont {{Sanchez-Puente}}},\ }\href {\doibase
  10.1088/1742-6596/600/1/012042} {\bibfield  {journal} {\bibinfo  {journal}
  {Journal of Physics Conference Series}\ }\textbf {\bibinfo {volume} {600}},\
  \bibinfo {eid} {012042} (\bibinfo {year} {2015})},\ \Eprint
  {http://arxiv.org/abs/1506.02145} {arXiv:1506.02145 [gr-qc]} \BibitemShut
  {NoStop}%
\bibitem [{\citenamefont {{Bazeia}}\ \emph {et~al.}(2014)\citenamefont
  {{Bazeia}}, \citenamefont {{Losano}}, \citenamefont {{Menezes}},
  \citenamefont {{Olmo}},\ and\ \citenamefont {{Rubiera-Garcia}}}]{palatini2}%
  \BibitemOpen
  \bibfield  {author} {\bibinfo {author} {\bibfnamefont {D.}~\bibnamefont
  {{Bazeia}}}, \bibinfo {author} {\bibfnamefont {L.}~\bibnamefont {{Losano}}},
  \bibinfo {author} {\bibfnamefont {R.}~\bibnamefont {{Menezes}}}, \bibinfo
  {author} {\bibfnamefont {G.~J.}\ \bibnamefont {{Olmo}}}, \ and\ \bibinfo
  {author} {\bibfnamefont {D.}~\bibnamefont {{Rubiera-Garcia}}},\ }\href@noop
  {} {\bibfield  {journal} {\bibinfo  {journal} {ArXiv e-prints}\ } (\bibinfo
  {year} {2014})},\ \Eprint {http://arxiv.org/abs/1411.0897} {arXiv:1411.0897
  [hep-th]} \BibitemShut {NoStop}%
\bibitem [{\citenamefont {{Lobo}}\ \emph {et~al.}(2014)\citenamefont {{Lobo}},
  \citenamefont {{Martinez-Asencio}}, \citenamefont {{Olmo}},\ and\
  \citenamefont {{Rubiera-Garcia}}}]{palatini3}%
  \BibitemOpen
  \bibfield  {author} {\bibinfo {author} {\bibfnamefont {F.~S.~N.}\
  \bibnamefont {{Lobo}}}, \bibinfo {author} {\bibfnamefont {J.}~\bibnamefont
  {{Martinez-Asencio}}}, \bibinfo {author} {\bibfnamefont {G.~J.}\ \bibnamefont
  {{Olmo}}}, \ and\ \bibinfo {author} {\bibfnamefont {D.}~\bibnamefont
  {{Rubiera-Garcia}}},\ }\href {\doibase 10.1103/PhysRevD.90.024033} {\bibfield
   {journal} {\bibinfo  {journal} {\prd}\ }\textbf {\bibinfo {volume} {90}},\
  \bibinfo {eid} {024033} (\bibinfo {year} {2014})},\ \Eprint
  {http://arxiv.org/abs/1403.0105} {arXiv:1403.0105 [hep-th]} \BibitemShut
  {NoStop}%
\bibitem [{\citenamefont {{Capozziello}}\ and\ \citenamefont
  {{Fang}}(2002)}]{SalvatoreQ1}%
  \BibitemOpen
  \bibfield  {author} {\bibinfo {author} {\bibfnamefont {S.}~\bibnamefont
  {{Capozziello}}}\ and\ \bibinfo {author} {\bibfnamefont {L.~Z.}\ \bibnamefont
  {{Fang}}},\ }\href {\doibase 10.1142/S0218271802002025} {\bibfield  {journal}
  {\bibinfo  {journal} {International Journal of Modern Physics D}\ }\textbf
  {\bibinfo {volume} {11}},\ \bibinfo {pages} {483} (\bibinfo {year} {2002})},\
  \Eprint {http://arxiv.org/abs/gr-qc/0201033} {gr-qc/0201033} \BibitemShut
  {NoStop}%
\bibitem [{\citenamefont {{Capozziello}}\ and\ \citenamefont
  {{DeFelice}}(2008)}]{SalvatoreQ2}%
  \BibitemOpen
  \bibfield  {author} {\bibinfo {author} {\bibfnamefont {S.}~\bibnamefont
  {{Capozziello}}}\ and\ \bibinfo {author} {\bibfnamefont {A.}~\bibnamefont
  {{DeFelice}}},\ }\href {\doibase 10.1088/1475-7516/2008/08/016} {\bibfield
  {journal} {\bibinfo  {journal} {\jcap}\ }\textbf {\bibinfo {volume} {8}},\
  \bibinfo {eid} {016} (\bibinfo {year} {2008})},\ \Eprint
  {http://arxiv.org/abs/0804.2163} {arXiv:0804.2163 [gr-qc]} \BibitemShut
  {NoStop}%
\end{thebibliography}%
